\documentclass[12pt]{article}
\usepackage[pdftex]{pict2e}
\usepackage{placeins}

\newcommand{\sclcar}{\mathfrak{scalcarr}}
\newcommand{\infcar}{\mathfrak{\Widetilde{confcarr}}}
\newcommand{\ccar}{\mathfrak{confcarr}}
\newcommand{\cca}{\mathfrak{cca}}

\pdfoutput=1
\usepackage{amsthm,amsmath,latexsym,amssymb,amsfonts,amscd,mathtools}
\usepackage{graphics,lscape,fancyhdr,array,stmaryrd,euscript}
\usepackage{ifthen,empheq,slashed}
\usepackage{color}
\numberwithin{equation}{section}
\usepackage{setspace}
\usepackage{cite}
\usepackage{mathrsfs}
\usepackage{amsmath}
\usepackage{cancel}

\usepackage[table]{xcolor}

\newcommand\Xcancel[2][black]{\renewcommand\CancelColor{\color{#1}}\xcancel{#2}}
\usepackage[hang,flushmargin]{footmisc} 
\usepackage[colorlinks,
			breaklinks,
			hyperfootnotes,
			linktoc=page,
            linkcolor=blue,
            citecolor=purple,
            urlcolor=blue,
            unicode,
            pdfstartview=FitH,
            bookmarks,
            bookmarksnumbered]{hyperref}

\usepackage{makecell}
\usepackage{geometry}

\newgeometry{left=0.8in,right=0.8in,top=.8in,bottom=1.1in}
\usepackage[most]{tcolorbox}
\usepackage{enumitem}
\usepackage{footnotebackref}
\usepackage{tablefootnote}

\usepackage{float}
\usepackage{stackengine}
\stackMath
\usepackage[retainorgcmds]{IEEEtrantools}
\usepackage{microtype}

\usepackage{tikz,pgfplots}
\pgfplotsset{compat=1.16}
\usetikzlibrary{calc,arrows,shapes,cd}
\usepackage{chemarrow}

\usepackage{tabularx}
\usepackage{caption}

\newcommand\boxedB[1]{{\setlength\fboxsep{8pt}\boxed{#1}}}

\DeclareFontFamily{U}{mathx}{\hyphenchar\font45 }
\DeclareFontShape{U}{mathx}{m}{n}{
	<-> mathx10
}{}
\DeclareSymbolFont{mathx}{U}{mathx}{m}{n}
\DeclareMathAccent{\Widetilde}{0}{mathx}{"72}

\def\be {\begin{equation}}
	\def\ee {\end{equation}}
\def\bea {\begin{eqnarray}}
	\def\eea {\end{eqnarray}}
\def\bc {\begin{center}}
	\def\ec {\end{center}}
\def\bg {\begin{align}}
	\def\eg {\end{align}}
\def\bi {\begin{itemize}}
	\def\ei {\end{itemize}}
\def\nn {\nonumber}

\def\le {\left}
\def\ri {\right}
\def\p {\partial}

\def\d  {\delta}

\def\beak{\begin{IEEEeqnarray*}}
\def\eeak{\end{IEEEeqnarray*}}
\def\nk{\IEEEyesnumber\phantomsection}

\begin{document}

 $\vspace{.5cm}$
	\begin{center}
		
{\bfseries \LARGE{Classification of Conformal Carroll Algebras}} \\
\par\noindent\rule{485pt}{0.8pt}		
\vskip 0.03\textheight
  Hamid \textsc{Afshar}\,\,$^{{\color{purple}a}}$, 
  Xavier \textsc{Bekaert}\,\,$^{{\color{purple}b}}$,
  Mojtaba \textsc{Najafizadeh}\,\,$^{{\color{purple}a},\, {\color{purple}c}}$ 	
		

{\raggedright 
 
\vspace*{15pt}
${}^{\color{purple}a}$~{ Department of Physics, Faculty of Science, Ferdowsi University of Mashhad, \\
    ~~~P.O.Box 1436, Mashhad, Iran }  		

 \vspace*{15pt}
${}^{\color{purple}b}$~{ Institut Denis Poisson, Unit\'e Mixte de Recherche 7013,\\
     ~~~Universit\'e de Tours -- Universit\'e d'Orl\'eans -- CNRS,\\
     ~~~Parc de Grandmont, 37200 Tours, France} 

 \vspace*{15pt}
 ${}^{\color{purple}c}$~{ School of Physics, Institute for Research in Fundamental Sciences (IPM), \\
      ~~~P.O.Box 19395-5531, Tehran, Iran }

   }

\vskip 0.02\textheight
		
{\tt\small
\href{mailto:ham.afshar@gmail.com}{ham.afshar@gmail.com},~
\href{mailto:xavier.bekaert@lmpt.univ-tours.fr}{xavier.bekaert@univ-tours.fr},~
\href{mailto:mnajafizadeh@ipm.ir}{mnajafizadeh@ipm.ir}
}

\vskip 0.05\textheight

{\sc\large Abstract} \end{center}

\vskip 0.02\textheight
  
\noindent  We classify a one-parameter family, $\mathfrak{confcarr}_z(d+1)$, of conformal extensions of the Carroll algebra in arbitrary dimension  with $z$ being the anisotropic scaling exponent. We further obtain their  infinite-dimensional extensions, $\Widetilde{\mathfrak{confcarr}}_z(d+1)$, and discuss their corresponding finite-dimensional truncated subalgebras when the scaling exponent is integer or half-integer. For all these conformal extensions, we also constrain the 2-point and 3-point correlation functions with electric  and/or magnetic features. 

\vskip 0.03\textheight
	

	\newpage
	\tableofcontents
	\newpage

\section{Introduction}
\begin{figure}[h!]
    \centering
    \begin{tikzpicture}[scale=4]
    \draw[->] (0,0,0) -- (0,0,1.8) node[below left] {~$D_t=a\,t\,\p_t$};
    \draw[->] (0,0,0) -- (2,0,0) node[right] {~$D_{x}=b\,x^i\p_i$};
    \draw[->] (0,0,0) -- (0,1.2,0) node[above] {~$K=c\,x^2\,\p_t$};

    \coordinate (A) at (0,0,0);
    \coordinate (B) at (1.5,0,0);
    \coordinate (C) at (1.5,.8,0);
    \coordinate (D) at (0,.8,0);
    \coordinate (E) at (0,0,1.2);
    \coordinate (F) at (1.5,0,1.2);
    \coordinate (G) at (1.5,.8,1.2);
    \coordinate (H) at (0,.8,1.2);
    
    \draw[purple] (A) -- (B) -- (C) -- (D) -- cycle;
    \draw[purple] (A) -- (E);
    \draw[purple] (B) -- (F);
    \draw[purple] (C) -- (G);
    \draw[purple] (D) -- (H);
    \draw[purple] (E) -- (F) -- (G) -- (H) -- cycle;

\draw[->, blue, dashed, out=45, in=180, looseness=1] (0, .8, 0) to node[left] {} (2.2, 1.4, 0.6) node[right] {\text{\hyperref[tK]{Type K algebra}}};

    \fill[green, opacity=0.4] (H) -- (G) -- (C) -- (D) -- cycle;
\draw[->, blue, dashed, out=45, in=180, looseness=2] (.75, .8, .6) to node[right] {} (2.2, .7, 0.6) node[right] {\text{\hyperref[tDK]{Type D-K algebras}}};

    \fill[pink, opacity=0.4] (A) -- (B) -- (F) -- (E) -- cycle;
\draw[->, blue, dashed, out=-45, in=180, looseness=2] (.75, 0, .6) to node[right] {} (2.2, -.5, 0.6) node[right] {\text{\hyperref[tD]{Type D algebras}}};

    \node[above right] at (A) {\text{\color{blue} \hyperref[Carroll]{Carroll}}};
    \node[below right] at (B) {$(0,b,0)$};
    \node[above right] at (C) {~$(0,b,c)$};
    \node[above left] at (D) {$(0,0,c)$};
    \node[left] at (E) {$(a,0,0)$~~};
    \node[below right] at (F) {$(a,b,0)$};     
    \node[right] at (G) {~$(a,b,c)$};
    \node[left] at (H) {$(a,0,c)$~~};

    \foreach \point in {A,B,C,D,E,F,G,H}
        \fill (\point) circle (.8pt);

\end{tikzpicture}
\caption{This figure illustrates the possible extensions of the Carroll algebra (located at the origin) with only the $K$ and $D\,(= D_t+ D_x)$ generators (i.e. $K_i=0$). The type K algebra is located on the K axis, and by rescaling the K generator, it can be positioned at a point on this axis. The type D algebras reside on the $D_t - D_x$ plane (pink plane). The type D-K algebras occupy the $D_t - D_x - K$ volume, which, through rescaling of the K generator, can be situated on a plane (green plane). Specifically, when $a,b,c$ are either $0$ or $1$, the special cases corresponding to the seven corners of the cube are listed in Table \ref{tabcc}.} 
\label{figcca}
\end{figure}
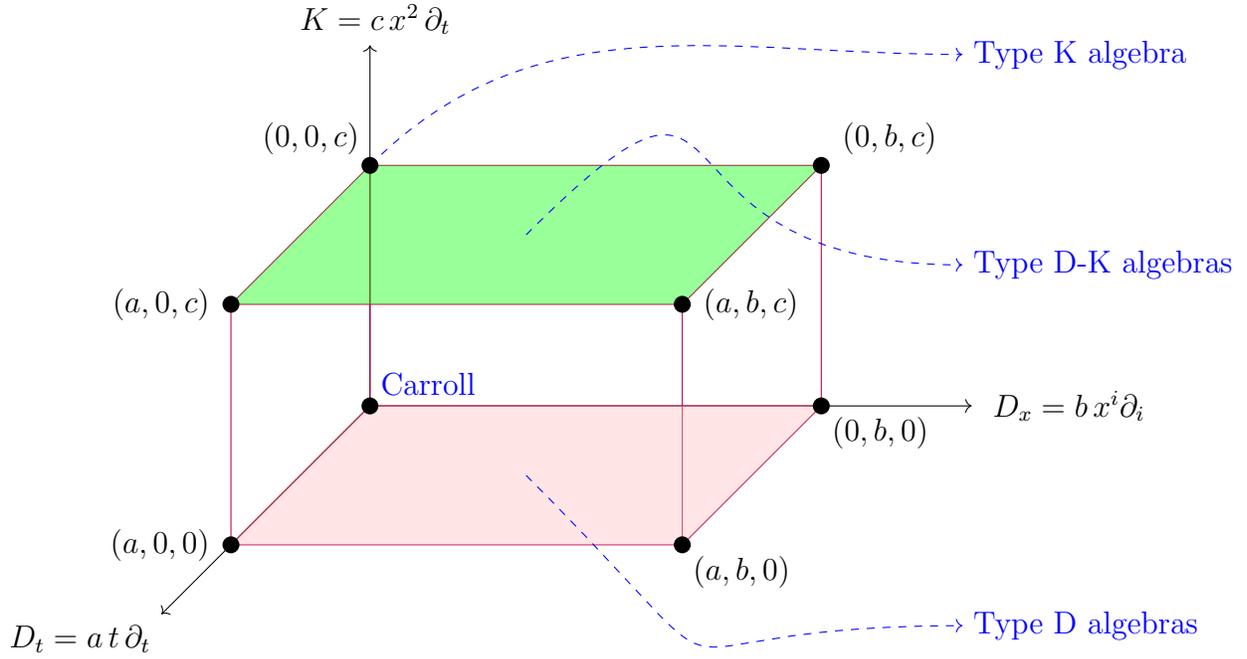

The Carroll symmetry \cite{Levy1965,SenGupta:1966qer,Bacry:1968zf} apart from possessing the usual 
time and space translations, as well as spatial rotations, is characterized by a specific boost-like symmetry;
\begin{align}
    \text{boost }(B_i):\qquad t\to t+\vec v\cdot \vec x\,,\qquad\vec x\to \vec x\,. 
\end{align}
This asymmetry between time and space already suggests the possibility of an anisotropic behaviour for time and space under a generic scaling;
\begin{align}
\text{dilatation }(D):\qquad t\to\lambda^z t\,,\qquad\vec x\to\lambda \,\vec x\,. 
\end{align}
We investigate the Carrollian enhancements of this anisotropic behaviour by applying  the closure condition on all possible conformal extensions of the Carroll algebra by the dilatation generator $D= D_t+ D_{x}$ (with $D_t=a\,t\,\partial_t$ and $D_x=b\,x^i\partial_i$) and by the generators $K$ and $K_i$ of ultra-relativistic special conformal transformations (SCT) whose finite forms are (for $a=z$ and $b=1$)
\begin{align}
&\text{temporal SCT }(K): \qquad t\to t+\alpha\,|\vec x|^2\,,  \qquad\vec x\to \vec x\,,\nn\\
&\text{spatial SCT }(K_i): \qquad t\to\frac{t}{\big(1-2\,\vec \beta\cdot\vec x+|\vec \beta|^2|\vec x|^2\big)^z}\,,  \qquad\vec x\to \frac{\vec x-|\vec x|^2\,\vec \beta}{1-2\,\vec \beta\cdot\vec x+|\vec \beta|^2|\vec x|^2} \,.
\end{align}
By themselves, these transformations only close if $z=1$.
In summary, our results are that, aside from the well-documented Carrollian conformal algebra (the ultra-relativistic contraction of the usual conformal algebra) in the literature\footnote{We use the word CCA for the ultra-relativistic contraction of the conformal algebra (Carrollian conformal algebra) not to be confused with conformal Carroll algebra as we obviously need to distinguish between the two in our work.}, the Carroll algebra can be conformally extended as illustrated in  Fig. \ref{figcca}. In this figure the Carroll algebra is located at the origin and possible extensions of it via $K$, $D_t$ and $D_x$ generators (with $K_i=0$) are illustrated along the three axis. The type K algebra is located on the $K$ axis, where upon rescaling, it can be positioned at a point on this axis. The type D algebras reside on the $D_t - D_x$ plane (pink plane). The type D-K algebras occupy the $D_t - D_x - K$ volume, which after rescaling the $K$ generator, can be situated on a plane (green plane). Specifically, when $a,b,c$ are either $0$ or $1$, the special cases corresponding to the seven corners of the cube are listed in Table \ref{tabcc} where the Carrollian conformal algebra is also represented for comparison. Setting $a=z$ and $b=c=1$, the type D-K algebras are denoted $\mathfrak{confcarr}_z(d+1)$. The inclusion of the spatial SCT generators $K_i$ is also discussed.

\FloatBarrier

\begin{table}[h!]
\begin{center}
\caption{ Different classes of minimal conformal extensions of the Carroll algebra \eqref{Carroll} corresponding to the corners of the cube in Figure \ref{figcca}, for $a,b,c=0,1$. Conformal extensions including the generators $K_i$ result in the Carrollian conformal algebra in any spacetime dimension and in the Carroll-Schr\"odinger algebra which is specific to $1+1$ dimension. The table displays only the non-trivial commutation relations.}
\begin{tabular}{ |p{4.5cm}|  }
\hline
 {\scriptsize \hyperref[tK]{\!\!\!\!\!\!\!Type K: (0,0,1)}} \beak{l}
[\,K\,,\,P_i\,]=-\,2\,B_i \vspace{-.4cm}
\eeak  
 \\
\hline
\end{tabular}\\
\begin{tabular}{ |p{4.5cm}|p{4.5cm}|p{4.5cm}|  }
\hline
 \cellcolor{pink!25} {\scriptsize \hyperref[tD]{\!\!\!\!\!\!\!Temporal scaling Carroll: (1,0,0)}}  \beak{l}
[\,D_t\,,\,H\,]=-\,H\\[0pt]
 [\,D_t\,,\,B_i\,]=-\,B_i \vspace{-.4cm}
 \eeak &  \cellcolor{pink!25}{\scriptsize \hyperref[tD]{\!\!\!\!\!\!\!Isotropic scaling Carroll: (1,1,0)}} \beak{l}
 [\,D\,,\,H\,]=-\,H \\[0pt]
 [\,D\,,\,P_i\,]=-\,P_i  \vspace{-.4cm}
\eeak &  \cellcolor{pink!25}{\scriptsize \hyperref[tD]{\!\!\!\!\!\!\!Spatial scaling Carroll: (0,1,0)}} \beak{l}
 [\,D_{x}\,,\,P_i\,]=-\,P_i\\[0pt]
 [\,D_{x}\,,\,B_i\,]=B_i \vspace{-.4cm}
\eeak \\
\hline
\end{tabular}  \\
\begin{tabular}{ |p{4.5cm}|p{4.5cm}|p{4.5cm}|  }
\hline
 \cellcolor{green!15}{\scriptsize \hyperref[tDK]{\!\!\!\!\!\!\!Temporal conformal Carroll: (1,0,1)}}
\beak{l}
 [\,D_t\,,\,H\,]=-\,H\\[0pt]
 [\,D_t\,,\,B_i\,]=-\,B_i \\[0pt]
 [\,D_t\,,\,K\,]=-\,K \\[0pt]
 [\,K\,,\,P_i\,]=-\,2\,B_i \vspace{-.4cm}
\eeak &  \cellcolor{green!15}{\scriptsize \hyperref[tDK]{\!\!\!\!\!\!\!Isotropic conformal Carroll: (1,1,1)}} 
\beak{l}
 [\,D\,,\,H\,]=-\,H\\[0pt]
 [\,D\,,\,P_i\,]=-\,P_i \\[0pt]
[\,D\,,\,K\,]=K \\[0pt]
 [\,K\,,\,P_i\,]=-\,2\,B_i \vspace{-.4cm}
\eeak &  \cellcolor{green!15}{\scriptsize \hyperref[tDK]{\!\!\!\!\!\!\!Spatial conformal Carroll: (0,1,1)}
}
\beak{l}
 [\,D_{x}\,,\,P_i\,]=-\,P_i\\[0pt]
[\,D_{x}\,,\,B_i\,]=B_i \\[0pt]
[\,D_{x}\,,\,K\,]=2\,K\\[0pt]
 [\,K\,,\,P_i\,]=-\,2\,B_i \vspace{-.4cm}
\eeak  \\
\hline
\end{tabular}
\\
\begin{tabular}{ |p{16.1cm}|  }
\hline
{\scriptsize \hyperref[cca]{\!\!\!\!\!\!\!Carrollian conformal algebra}}
\beak{lll}
[\,D\,,\,H\,]=-\,H      \qquad & [\,D\,,\,K_i\,]=K_i    \qquad & [\,K_i\,,\,P_j\,]=-\,2\,\d_{ij}\,D-\,2\,J_{ij} \\[0pt]
[\,D\,,\,P_i\,]=-\,P_i  \qquad & [\,K\,,\,P_i\,]=-\,2\,B_i  \qquad & [\,K_i\,,\,B_j\,]=-\,\d_{ij}\,K \\[0pt]
[\,D\,,\,K\,]=K         \qquad & [\,K_i\,,\,H\,]=-\,2\,B_i  \qquad & [\,K_i\,,\,J_{jk}\,]=\d_{i[j}K_{k]} \vspace{-.4cm}
\eeak \\
\hline
\end{tabular}
\begin{tabular}{ |p{16.1cm}|  }
\hline
{\scriptsize \hyperref[c-sch-a]{\!\!\!\!\!\!\!Carroll-Schr\"odinger algebra}}
\beak{lll}
[\,D\,,\,H\,]=-\,H      \qquad & [\,D\,,\,B\,]=B    \qquad & [\,\mathbb{K}\,,\,H\,]=-\,B \\[0pt]
[\,D\,,\,P\,]=-\,2\,P   \qquad & [\,D\,,\,\mathbb{K}\,]=2\,\mathbb{K}  \qquad & [\,\mathbb{K}\,,\,P\,]=-\,D \vspace{-.4cm}
\eeak \\
\hline
\end{tabular}
\end{center}
\label{tabcc}
\end{table}

\FloatBarrier

The Carroll algebra $\mathfrak{carr}(d+1)$ admits infinite-dimensional conformal extensions as well. The most famous one is denoted as $\mathfrak{bms}_{d+2}$, it is the  algebra of the BMS group of asymptotic symmetries for asymptotically flat $d+2$ dimensional spacetimes, first discovered in 1962 by Bondi, van der Burg, Metzner and Sachs \cite{Bondi:1962px,Sachs:1962zza} for $d=2$. This type of enhancement of the Carroll algebra was discussed in \cite{Duval:2014uva} and denoted $\mathfrak{ccarr}_2(d+1)$. It basically suggested that the holographic dual of gravity theories in asymptotically flat spacetime should be a conformal extension of Carrollian field theories (Carrollian CFT) in one lower dimension. 
This is well  manifested as the isomorphism between the asymptotic symmetry group of the flat space (BMS) and the Carrollian contraction of the conformal group in one-lower dimension \cite{Duval:2014uva},
\begin{align}\label{bmscca1}
\mathfrak{ccarr}_2(d+1) \cong \mathfrak{bms}_{d+2}\,.
\end{align}
Obviously, in the correspondence  \eqref{bmscca1} there is an important role played by the specific  boundary conditions imposed in flat spacetime \cite{Bondi:1962px,Sachs:1962zza,Barnich:2006av,Barnich:2009se,Barnich:2010eb,Troessaert:2017jcm,Henneaux:2018cst}. Given that one can consider different physically motivated consistent flat-space boundary conditions like Rindler or near horizon boundary conditions \cite{Afshar:2015wjm,Donnay:2015abr,Afshar:2016wfy,Donnay:2016ejv,Grumiller:2019ygj,Grumiller:2019fmp,Adami:2023wbe}, it is suggestive that the algebra $\mathfrak{bms}_{d+2}$  generalizes to different asymptotic symmetry algebras after imposing other locally flat boundary conditions. Algebraically, this leads to possible consistent infinite-dimensional conformal extensions of the Carroll group. In fact, it was shown  \cite{Duval:2014lpa} that the left-hand side of the above correspondence  generalizes to a discrete one-parameter family of infinite-dimensional algebras $\mathfrak{ccarr}_\ell(d+1)$ where the integer $\ell=0, 1, 2, 3, \ldots, \infty$ is  called the level. As will be emphasised in the present paper, from the present algebraic perspective there is actually no reason to restrict the level to be an integer rather than any real number. We will discuss the continuous one-parameter generalisation of these algebras, which will be indexed instead by the dynamical exponent $z$ and denoted $\Widetilde{\mathfrak{confcarr}}_z(d+1)$. In this sense, all discrete generalizations in \cite{Duval:2014lpa} embed into our infinite-dimensional conformal extensions of the Carroll algebra for $z=\frac{2}{\ell}=\infty, 2, 1, \frac23, \ldots, 0$. The fact that in our case $z$ is a free parameter will have an important consequence because a consistent finite-dimensional truncation can occur when $z$ is an integer $N\geqslant1$ (for dimension $d\geqslant2$) and we denote these finite-dimensional algebras $\cca_{N}(d+1)$. A well-known case is $N=1$ where the truncated algebra is the Carrollian conformal algebra $\cca_1(d+1)\cong\mathfrak{iso}(d+1,1)$ with $d\geqslant2$, which is the In\"on\"u-Wigner contraction of the relativistic conformal algebra $\mathfrak{so}(d+1,2)$ in the ultra-relativistic limit. For $d=1$, i.e. in spacetime dimension $1+1$, there exist consistent truncations at integer and half-integer values $z=N/2\geqslant1/2$, where the case $z=1/2$ corresponds to the Carroll-Schr\"odinger algebra and the case $z=1$ corresponds to the 2D Carrollian conformal algebra.

After identifying consistent conformal extensions of the Carroll algebra, we study the corresponding conformally-invariant Carrollian field theories  (Carrollian CFT's)  by constraining the consistent two and three point functions between scalar primary fields.\footnote{A primary field in a physical system with Lifshitz invariance exhibits a general scaling dimension $\Delta$ defined according to its transformation under  scaling
$    \phi(t,\vec x)\to \lambda^{\Delta}\phi(\lambda^zt,\lambda \vec x)\,
$.} We show that a generic 2-point function consists of two pieces: a magnetic (time-independent) piece and an electric (ultra-local in space) piece. The 3-point functions exhibit the same general behaviour with possible mixed terms in which  distances  are collinear.

The layout of this paper is as follows. In Section \ref{classification}, we classify minimal conformal extensions of the Carroll algebra, identifying five types: four types in any spacetime dimension and one type specific to two dimensions. In Section \ref{infdim}, we present infinite-dimensional conformal extensions of the Carroll algebra. These are characterized by an arbitrary critical exponent $z$ in two, three, and higher spacetime dimensions. In Section \ref{wardidentity}, we study the two and three point functions invariant under various conformally extended symmetries.  In Section \ref{Carrollianinversion}, we obtain the finite form of the Carrollian SCT's which enables us to constrain the correlation functions directly from covariance under inversion and translation. Finally, in Section \ref{conclu}, we briefly summarize our findings and suggest a few directions for future research.

\section{Minimal conformal extensions of the Carroll algebra}\label{classification}

Let us consider the set $\{H, P_i, B_j, J_{ij}\}$  of generators of the Carroll algebra \cite{Levy1965,SenGupta:1966qer,Bacry:1968zf} (see also \cite{Figueroa-OFarrill:2017ycu,Figueroa-OFarrill:2017tcy}), i.e. the generators $P_i$ and $H$ of space and time translations respectively,  the generators $B_i$ of Carroll boosts and the generators $J_{ij}$ of rotations ($i,j=1,2,\dots,d$). 

One possible (infinite-dimensional) extension of the Carroll algebra $\mathfrak{carr}(d+1)$ is via supertranslation generators $M_f:=f(x^i)\,\partial_t$\,, of which the generators $H$ of time translations and $B_i$ of Carroll boosts correspond to the particular case when $f$ is an affine function of the spatial Cartesian coordinates. 
The infinite-dimensional extension of the Carroll algebra with generators $\{M_f,P_i,J_{ij}\}$
is the Lie algebra of Carrollian isometries of flat Carroll spacetime, which will be denoted $\mathfrak{isocarr}(d+1)$ and which has the structure of a semi-direct sum $\mathfrak{iso}(d)\inplus C^\infty(\mathbb{R}^d)$ where the Euclidean algebra $\mathfrak{iso}(d)$ generated by $\{P_i,J_{ij}\}$ acts on the abelian ideal $C^\infty(\mathbb{R}^d)$ of supertranslations. If the Euclidean algebra $\mathfrak{iso}(d)$ is extended to the conformal algebra $\mathfrak{so}(d+1,1)$ of the Euclidean space $\mathbb{R}^d$, then one gets the BMS algebra $\mathfrak{bms}_{d+2}$ \cite{Duval:2014uva}.  
These infinite-dimensional extensions are well-known but we would like to focus first on generic finite-dimensional conformal extensions.

On top of the kinematical generators of the Carroll algebra, let us introduce the following generators $\{D, K, K_i\}$ of conformal Carroll transformations with some arbitrary parameters $a,b,c,\alpha,\beta,\gamma$: 
\beak{lll}\phantomsection\label{CCG}
H=\p_t\,,\qquad\qquad & P_i=\p_i \,, \qquad\qquad &  B_i=x_i\,\p_t\,,\qquad\qquad  J_{ij} = x_i\,\p_j-x_j\,\p_i \\[5pt]
D={\color{blue}a}\,t\p_t+{\color{blue}b}\,x^i\p_i\,,\quad\quad & K={\color{blue}c}\,x^2\p_t\,,\quad\quad &
K_i=2\,{\color{blue}\alpha}\,x_i \,t\p_t+2\,{\color{blue}\beta}\,x_i\, x^j\p_j - {\color{blue}\gamma}\,x^2\p_i\,,\nk
\eeak
where $D$ generates dilatations, while $K_i$ and $K$ generate spatial and temporal special conformal transformations, respectively. Note that the temporal SCT generator $K$ is an example of supertranslation generator $M_f=f(x^i)\partial_t$ whose function $f$ is rotation-invariant and quadratic in the spatial coordinates $x^i$.
We assumed that these generators have the same type of dependence as in the usual case but, apart from that, we allowed for \textit{a priori} arbitrary anisotropy between space and time. It is understood that any commutator of the differential operators \eqref{CCG} automatically satisfies Jacobi identity as it defines a Lie bracket between vector fields, however the closure of the algebra should be checked as is illustrated below.

The generators of the Carroll algebra appear in the first line of \eqref{CCG}. Their commutation relations are
\cite{Levy1965,SenGupta:1966qer,Bacry:1968zf,Figueroa-OFarrill:2017ycu,Figueroa-OFarrill:2017tcy})
\beak{l} \phantomsection\label{Carroll}
[\,P_i\,,\,B_j\,]=\d_{ij}\,H\,, \\[0pt]
[\,P_i\,,\,J_{jk}\,]=\d_{i[j}\,P_{k]}\,,\\[0pt] 
[\,B_i\,,\,J_{jk}\,]=\d_{i[j}\,B_{k]}\,,\\[0pt]
[\,J_{ij}\,,\,J_{kl}\,]=\d_{[i[k}J_{l]j]}\,.\nk 
\eeak 
As one can see the Carroll algebra has the structure of a semi-direct sum $\mathfrak{carr}(d+1)=\mathfrak{so}(d)\inplus\mathfrak{h}_d$, of the Heisenberg algebra $\mathfrak{h}_d$ generated by the generators $\{H, P_i, B_j\}$, on which the ideal $\mathfrak{so}(d)$, spanned by the generators $\{J_{ij}\}$, acts via rotations of the indices.

Taking into account the conformal generators in the second line of \eqref{CCG}, one gets the following non-trivial commutation relations
\beak{ll} \phantomsection\label{comm}
[\,D\,,\,H\,]=-\,a\,H\,,\qquad\qquad & [\,K\,,\,K_i\,]=2\,x_i\,K\,(\alpha-2\,\beta+\gamma)\,, \\[2pt]
[\,D\,,\,P_i\,]=-\,b\,P_i\,,\qquad\qquad & [\,K_i\,,\,H\,]=-\,2\,\alpha\,B_i\,, \\[2pt] 
[\,D\,,\,B_i\,]=(b-a)\,B_i\,,\qquad\qquad &  [\,K_i\,,\,P_j\,]= -\,2\,\d_{ij}\,(\alpha\,t\,\p_t+\beta\,x^k\p_k)-J_{ij}\,(\beta+\gamma)-\tilde{J}_{ij}\,(\beta-\gamma)\,,\\[2pt]
[\,D\,,\,K\,]=(2b-a)\,K\,,\qquad\qquad &  [\,K_i\,,\,B_j\,]=-\,2\,x_i\,B_j\,(\alpha-\beta)-\,\d_{ij}\,\gamma\,x^2\,\p_t\,,  \\[2pt]
[\,D\,,\,K_i\,]=b\,K_i\,,\qquad\qquad & [\,K_i\,,\,J_{jk}\,]=\d_{i[j}K_{k]}\,, \\[2pt]
[\,K\,,\,P_i\,]=-\,2\,c\,B_i\,, \qquad\qquad & [\,K_i\,,\,K_j\,]=2\,\gamma\,(\gamma-\beta)\,x^2\,J_{ij}\,, \nk
\eeak
where $\tilde{J}_{ij}:=x_i\,\p_j\,+\,x_j\,\p_i$\,. These commutation relations reveal that additional generators have emerged on the right-hand side, specifically $x_i\,K$,\, $\tilde{J}_{ij}$,\, $x_i\,B_j$ and $x^2\,J_{ij}$, which are not among the main set \eqref{CCG}. To ensure that the closure is satisfied, the coefficients of these additional generators must be set to zero. By enforcing this condition, we can determine all possible values for the parameters $a$, $b$, $c$, $\alpha$, $\beta$, $\gamma$, which leads to the different types of conformal Carroll algebras.

In general, one can think of the Carroll algebra \eqref{Carroll} as being conformally extended by adding the conformal generators ($D, K, K_i$) in seven different ways, denoted by
\be 
D,\quad\qquad  K,\quad\qquad {\Xcancel[red]{K_i}},\quad\qquad D\!-\!K,\quad\qquad {\Xcancel[red]{D\!-\!K_i}}, \quad\qquad {\Xcancel[red]{K\!-\!K_i}}, \quad\qquad D\!-\!K\!-\!K_i\,. \label{7types}
\ee 
However, one immediately finds that satisfying the closure of commutation relations cancels some of the mentioned cases in \eqref{7types}, as indicated above.

For example, consider the case where we add just the generators $K_i$ to the Carroll algebra \eqref{Carroll}. In this case, we would have $a=b=c=0$, while $\alpha$, $\beta$, $\gamma$ are not all vanishing. However, in order to close the commutation relations, the commutator $[\,K_i\,,\,P_j\,]$ in \eqref{comm} demonstrates that $\alpha, \beta,\gamma$ must be set to zero, implying that $K_i=0$. However, this is inconsistent with the initial assumption that $K_i$ was non-zero. As a result, it is not possible to extend the Carroll algebra conformally by adding only the generators $K_i$, and this case was therefore cancelled in \eqref{7types}. Similarly, we can demonstrate that the extensions involving $D\!-\!K_i$ and $K\!-\!K_i$ are not possible as well.

The four remaining types in \eqref{7types} can be determined as follows. Depending on whether we have only dilatation $D$, only temporal special conformal transformation $K$, or both, we find three distinct types of algebras. Notably, these types are characterized by the absence of spatial SCT's, implying $K_i=0$ and consequently $\alpha=\beta=\gamma=0$. Therefore, we denote these algebras by $(a,b,0)$, $(0,0,c)$, and $(a,b,c)$, respectively. When $K_i$ is non-zero, a fourth type of algebra emerges, namely the Carrollian conformal algebra, which encompasses the generators $D$, $K$, and $K_i$. Therefore, the four general types of conformal Carroll algebras, derived and explained in this section, are:
\begin{equation*}
	\begin{cases}
            ~\text{Type $K$ algebra} & \text{or \quad the algebra $(0, 0, c)$; \qquad\qquad\qquad\qquad\qquad\, Subsec. \ref{tyK} }\,, \\[5pt]
		~\text{Type $D$ algebras} & \text{or \quad the algebra $(a, b, 0)$; \qquad\qquad\qquad\qquad\qquad\, Subsec. \ref{tyD} }\,, \\[5pt]
            ~\text{Type $D\!-\!K$ algebras} & \text{or \quad the algebra $(a, b, c)$; \qquad\qquad\qquad\qquad\qquad\, Subsec. \ref{tyDK} }\,, \\[5pt]
            ~\text{Type $D\!-\!K\!-\!K_i$ algebra} & \text{or \quad the Carrollian conformal algebra; ~~\,\,\quad~\quad\,\quad Subsec. \ref{tyDKKi} }\,.
	\end{cases}
\end{equation*}

It is worth noting that the $D\!-\!K_i$ case, which was previously ruled out in \eqref{7types}, can in fact be a possible type of algebra in $1+1$ dimensions. This type with central charge corresponds to the ``Carroll-Schr\"odinger algebra'' identified in \cite{Najafizadeh:2024imn}. Therefore, this is the fifth type of conformal Carroll algebras, which we will reproduce in Section \ref{csa}.


 \subsection{Type K algebra: \texorpdfstring{$\mathfrak{Kcarr}(d+1)$}{} } \label{tyK}
 

This type of algebra corresponds to the case \texorpdfstring{$(0,0,c)$}{} with $c\neq 0$, which we set to $c=1$ without loss of generality. In this case, the Carroll algebra is extended only by the temporal special conformal transformation generator $K$. Therefore, in \eqref{CCG}, the parameters $a$, $b$, $\alpha$, $\beta$, $\gamma$ are all zero, resulting in the generators $D$ and $K_i$ to be trivially realised. The nontrivial generators are
\beak{lll}\phantomsection
H=\p_t\,,\qquad\qquad & P_i=\p_i \,, \qquad\qquad &  B_i=x_i\,\p_t\,,\qquad\qquad  J_{ij} = x_i\,\p_j-x_j\,\p_i\,, \\[5pt]
K=x^2 \p_t\,. \label{tsct}\nk
\eeak
These generators \eqref{tsct} satisfy the following non-zero commutation relations 
\beak{ll}\phantomsection\label{tK}
[\,P_i\,,\,B_j\,]=\d_{ij}\,H\,,\qquad &  [\,K\,,\,P_i\,]=-\,2\,B_i\,,\\[0pt]
[\,P_i\,,\,J_{jk}\,]=\d_{i[j}\,P_{k]}\,,\qquad & \\[0pt]
[\,B_i\,,\,J_{jk}\,]=\d_{i[j}\,B_{k]}\,,\qquad & \\[0pt]
[\,J_{ij}\,,\,J_{kl}\,]=\d_{[i[k}J_{l]j]}\,,\nk
\eeak
which can also be obtained directly from \eqref{comm} by setting the parameters $a$, $b$, $\alpha$, $\beta$, $\gamma$ to zero, and $c=1$. We will refer to the algebra \eqref{tK} as the ``conformal Carroll algebra of type K'' or the ``temporal SCT-Carroll algebra'' and denote it $\mathfrak{Kcarr}(d+1)$. Note that the derived algebra\footnote{The \textit{derived algebra} $\mathfrak{g}':=[\mathfrak{g},\mathfrak{g}]$ of a Lie algebra $\mathfrak{g}$ is the ideal spanned by elements which can be written as sums of Lie brackets of elements of $\mathfrak{g}$.} of this type K conformal Carroll algebra is nothing but the Carroll algebra itself, $\mathfrak{Kcarr}(d+1)'=\mathfrak{carr}(d+1)$, since the generator $K$ is the only one that does not appear on the right-hand side of the commutators in \eqref{tK}.

We note that the algebra \eqref{tK} is the most basic conformal extension of the Carroll algebra, which stands in contrast to the Poincar\'e algebra. Notably, the latter cannot be extended solely by incorporating the temporal special conformal transformation $K$ in the relativistic case (because of Lorentz covariance), a possibility that is afforded by the Carroll algebra.

 \subsection{Type D algebras: \texorpdfstring{$\mathfrak{scalcarr}_z(d+1)$}{} }\label{tyD}
This type of algebras corresponds to the case \texorpdfstring{$(a,b,0)$}{} where $a$ and $b$ cannot both vanish. In this case, the Carroll algebra is extended only by the dilatation generator $D$. Thus, in \eqref{CCG}, the parameters $c$, $\alpha$, $\beta$, $\gamma$ are all zero, resulting in the generators $K$ and $K_i$ to be trivially realised. The nontrivial generators are
\beak{lll}\phantomsection\label{tdgen}
H=\p_t\,,\qquad\qquad & P_i=\p_i \,, \qquad\qquad &  B_i=x_i\,\p_t\,,\qquad\qquad  J_{ij} = x_i\,\p_j-x_j\,\p_i \\[5pt]
D=a\,t\,\p_t+b\,x^i\p_i\,, \nk
\eeak
where $a,b$ are arbitrary. These generators satisfy the following non-zero commutation relations
\beak{ll}\phantomsection\label{tD}
[\,P_i\,,\,B_j\,]=\d_{ij}\,H\,,\qquad &  [\,D\,,\,H\,]=-\,a\,H\,,\\[0pt]
[\,P_i\,,\,J_{jk}\,]=\d_{i[j}\,P_{k]}\,,\qquad & [\,D\,,\,P_i\,]=-\,b\,P_i\,,\\[0pt]
[\,B_i\,,\,J_{jk}\,]=\d_{i[j}\,B_{k]}\,,\qquad & [\,D\,,\,B_i\,]=(b-a)\,B_i\,,\\[0pt]
[\,J_{ij}\,,\,J_{kl}\,]=\d_{[i[k}J_{l]j]}\,, \nk
\eeak
which can also be obtained directly from \eqref{comm} by setting the parameters $c$, $\alpha$, $\beta$, $\gamma$ to zero. We refer to the algebras in \eqref{tD} as the ``conformal Carroll algebras of type D'' or ``scaling Carroll algebras''.\footnote{This algebra was called ``boost-extended Lifshitz algebra'' in Appendix B of \cite{Gibbons:2009me}, and was called ``Lifshitz Carroll algebra'' in \cite{Bergshoeff:2015wma}. Our choice of terminology is inspired from the name ``scaling Poincar\'e algebra'' in \cite{Nakayama:2023xzu} for its relativistic analogue.} When the parameters $a$ and $b$ take specific values, the algebra \eqref{tD} gives rise to special cases, including temporal, isotropic, and spatial scaling Carroll algebras, as listed in Table \ref{tabcc}.  

We note that, for $b\neq 0$, by rescaling in \eqref{tdgen} the dilatation generator $D\to \tilde{D}=({a}/{b})\,t\,\p_t+x^i\p_i$ and introducing the new parameter $z=a/b$, the algebra \eqref{tD} can obviously be expressed in terms of a one-parameter family instead of a fake two-parameter one. The case $b=0$ can even be accounted by formally allowing the value $z=\infty$ in the family of algebras. In particular, to identify finite-dimensional algebras, we will sometimes use the two-parameter ($a, b$) family, while the one-parameter ($z$) family can be more suitable for illustrating infinite-dimensional algebras. 

In conclusion, there is a one-parameter family of scaling Carroll algebras $\sclcar_z(d+1)$ indexed by the parameter $z$ which is the analogue of the dynamical exponent $z$ in the Lifshitz algebra. They can be readily expressed by setting $a=z$ and $b=1$ in \eqref{tdgen}, \eqref{tD}. In particular, the dilatation generator takes the form $D=z\,t\,\p_t+x^i\p_i$ for $z\in\mathbb R$ and $D=t\,\p_t$ for $z=\infty$. 
Accordingly, the algebra $\mathfrak{scalcarr}_0(d+1)$ will be called the ``spatial scaling Carroll algebra'' while the algebra $\sclcar_\infty(d+1)$ will be called the ``temporal scaling Carroll algebra''. Finally, the algebra $\sclcar_1(d+1)$ will be called the ``isotropic scaling Carroll algebra'' since for $z=1$ the scaling is the same in space and time, like in the relativistic case.

Note that the derived algebra of any scaling Carroll algebra is nothing but the Carroll algebra itself since the dilatation generator $D$ is the only one that does not appear on the right-hand side of the commutators in \eqref{tD}, i.e. $\sclcar_z(d+1)'=\mathfrak{carr}(d+1)$ for any $z$.

 \subsection{Type D-K algebras: \texorpdfstring{$\mathfrak{confcarr}_z(d+1)$}{} } \label{tyDK}

This type of algebras corresponds to the case \texorpdfstring{$(a,b,c)$}{} with $ab\neq 0$ and $c\neq 0$. In this case, the conformal extension of the Carroll algebra incorporates both the generator $D$ of dilatations and the generator $K$ of temporal special conformal transformations. In \eqref{CCG}, the parameters $\alpha$, $\beta$, $\gamma$ are all zero, leading to the vanishing of the $K_i$ generator. The nontrivial generators are then
\beak{lll}\phantomsection\label{tDKgene}
H=\p_t\,,\qquad\qquad & P_i=\p_i \,, \qquad\qquad &  B_i=x_i\,\p_t\,,\qquad\qquad  J_{ij} = x_i\,\p_j-x_j\,\p_i \\[5pt]
D=a\,t\,\p_t+b\,x^i\p_i\,,\quad\quad & K=x^2\p_t\,,\nk
\eeak
where the generator $K$ is rescaled by a factor of $c$ (or, equivalently, setting $c=1$). These generators satisfy the following non-zero commutation relations
\beak{lll}\phantomsection\label{tDK}
[\,P_i\,,\,B_j\,]=\d_{ij}\,H\,,\qquad\quad &  [\,D\,,\,H\,]=-\,a\,H\,, \qquad\quad &  [\,K\,,\,P_i\,]=-\,2\,B_i\,,\\[0pt]
[\,P_i\,,\,J_{jk}\,]=\d_{i[j}\,P_{k]}\,,\qquad\quad & [\,D\,,\,P_i\,]=-\,b\,P_i\,, \\[0pt]
[\,B_i\,,\,J_{jk}\,]=\d_{i[j}\,B_{k]}\,,\qquad\quad & [\,D\,,\,B_i\,]=(b-a)\,B_i\,, \\[0pt]
[\,J_{ij}\,,\,J_{kl}\,]=\d_{[i[k}J_{l]j]}\,, \qquad\quad & [\,D\,,\,K\,]=(2b-a)\,K\,,\nk
\eeak
which again can be directly obtained from \eqref{comm} by setting $\alpha=\beta=\gamma=0$ and $c=1$. We refer to the algebras in \eqref{tDK} as the ``conformal Carroll algebras of type D-K''
or ``conformal Carroll algebras of dynamical exponent $z$''. We will denote them $\ccar_z(d+1)$. For $z=0$, it will be called ``spatial conformal Carroll algebra'' $\ccar_0(d+1)$. Note that it
 corresponds to the algebra discussed in Section III.D of \cite{Duval:2014lpa}, where it was denoted $\mathfrak{schcarr}(d+1)$
because it is a subalgebra of the corresponding Schr\"odinger algebra $\mathfrak{sch}(d+1)$.
 The case $z=\infty$ will be called the ``temporal conformal Carroll algebra''  while the case $z=1$ will be called the ``isotropic conformal Carroll algebra''. Note that, for any $z$, we have the following hierarchy of Lie subalgebras:
\begin{equation}\label{hierarchy}
\mathfrak{carr}(d+1)\subset\sclcar_z(d+1)\subset\ccar_z(d+1)\,.
\end{equation}
Moreover, let us mention that the derived algebra of the conformal Carroll algebra of type D-K  is the conformal Carroll algebra of type K, i.e. $\ccar_z(d+1)'=\mathfrak{Kcarr}(d+1)$,  since the generator $K$ appears on the right-hand side of the last commutator in \eqref{tDK} but not the generator $D$.

\subsection{Type D-K-K\texorpdfstring{$_i$}{} algebra: \texorpdfstring{$\cca_1(d+1)$}{} } \label{tyDKKi}

This type of algebra extends the Carroll algebra in a conformal manner by including all additional generators.
Therefore, in \eqref{CCG} and \eqref{comm}, all parameters $a$, $b$, $c$, $\alpha$, $\beta$, $\gamma$ are \textit{a priori} non-zero. We can determine the values of these parameters as follows.

By setting the coefficients of the terms $\tilde{J}_{ij}$ and $x_i\,B_j$ to zero on the right-hand side of the commutation relations in \eqref{comm}, we find that $\alpha=\beta=\gamma$. In addition, the commutator of $[\,K_i\,,\,P_j\,]$ leads to the choice $a=\alpha$ and $b=\beta$, while the commutator of $[\,K_i\,,\,B_j\,]$ leads to choosing $c=\gamma$. Therefore, if the Carroll algebra is conformally extended by $D$, $K$, and $K_i$, then we find that all the parameters must be equal. Strictly speaking, there is a loophole in this argument when $d=1$ because, in this degenerate case, the index $i$ is somewhat meaningless, so the terms with coefficient $\beta$ and $\gamma$ are of the same type. In fact, in spacetime dimension 1+1, there is an extra conformal extension, that will be discussed in Section \ref{csa}.

In the generic case, the parameters must all be equal, so they can be absorbed into the generators by rescaling, or equivalently, the parameters can simply be set to one: $a=b=c=\alpha=\beta=\gamma=1$. As a result of this, the generators in \eqref{CCG} become
\beak{lll}\phantomsection\label{ccagen}
H=\p_t\,,\quad\quad & P_i=\p_i \,, \quad\quad &  B_i=x_i\,\p_t\,,\qquad\qquad  J_{ij} = x_i\,\p_j-x_j\,\p_i\,, \\[5pt]
D=t\p_t+x^i\p_i\,,\quad\quad & K=x^2\p_t\,,\quad\quad &
K_i=2\,x_i\,(t\p_t+x^j\p_j)-\,x^2\p_i\,, \nk
\eeak 
and the commutation relations in \eqref{comm} reduce to the following non-zero brackets
\beak{lll}\phantomsection\label{cca}
[\,P_i\,,\,B_j\,]=\d_{ij}\,H\,,\qquad & [\,D\,,\,H\,]=-\,H\,,     \qquad & [\,K\,,\,P_i\,]=-\,2\,B_i \,, \\[0pt]
[\,P_i\,,\,J_{jk}\,]=\d_{i[j}\,P_{k]}\,, \qquad & [\,D\,,\,P_i\,]=-\,P_i\,, \qquad & [\,K_i\,,\,H\,]=-\,2\,B_i\,, \\[0pt]
[\,B_i\,,\,J_{jk}\,]=\d_{i[j}\,B_{k]}\,, \qquad & [\,D\,,\,K\,]=K\,, \qquad &  [\,K_i\,,\,P_j\,]=-\,2\,\d_{ij}\,D-\,2\,J_{ij}\,,  \\[0pt]
[\,J_{ij}\,,\,J_{kl}\,]=\d_{[i[k}J_{l]j]}\,, \qquad & [\,D\,,\,K_i\,]=K_i\,,\qquad & [\,K_i\,,\,B_j\,]=-\,\d_{ij}\,K\,, \\[0pt]
& & [\,K_i\,,\,J_{jk}\,]=\d_{i[j}K_{k]}\,. \nk
\eeak
In other words, we recovered the Carrollian conformal algebra. It is isomorphic to $\mathfrak{iso}(d+1,1)$.
Note that we have the following hierarchy of Lie subalgebras:
\begin{equation}\label{bmshierarch}
\mathfrak{carr}(d+1)\subset\sclcar_1(d+1)\subset\ccar_1(d+1)\subset\cca_1(d+1)\cong\mathfrak{iso}(d+1,1)\subset\mathfrak{bms}_{d+2}\,.
\end{equation}
Moreover, the Carrollian conformal algebra is a so-called ``perfect'' Lie algebra, i.e. it is equal to its derived Lie algebra: $\mathfrak{iso}(d+1,1)'=\mathfrak{iso}(d+1,1)$ since all generators appear on the right-hand side of \eqref{cca}.

\subsection{Type D-K\texorpdfstring{$_i$}{} algebra: \texorpdfstring{$\mathfrak{carrsch}(1+1)$}{} } \label{csa}


In $1+1$ spacetime dimensions, the rotation generator $J_{ij}$ vanishes, and the conformal Carroll algebra generators \eqref{CCG} reduce to
\beak{lll}\phantomsection \label{cca1+1}
H=\p_t\,,\qquad\qquad\quad  & P=\p_x \,, \qquad\qquad\quad  &  B=x\,\p_t\,,  \nk\\[5pt]
D={\color{blue}a}\,t\,\p_t+{\color{blue}b}\,x\,\p_x\,,\quad \qquad \qquad  & K={\color{blue}c}\,x^2\p_t\,,\quad \qquad \qquad  &
\mathbb{K}=2\,{\color{blue}\alpha}\,x \,t\p_t+{\color{blue}\delta}\,x^2\,\p_x \,,
\eeak
where $\delta:=2\,\beta-\gamma$ is a new parameter, and $\mathbb{K}$ denotes the spatial special conformal transformation $K_i$ in one spatial dimension. The generators \eqref{cca1+1} satisfy the following non-zero commutation relations
\beak{ll}\phantomsection \label{1+1com}
[\,P\,,\,B\,]=H\,,        \qquad & [\,K\,,\,P\,]=-\,2\,c\,B\,, \\[0pt]
[\,D\,,\,H\,]=-\,a\,H\,,  \qquad & [\,K\,,\,\mathbb{K}\,]=-\,2\,x\,K\,(\delta-\alpha)\,, \\[0pt]
[\,D\,,\,P\,]=-\,b\,P\,,  \qquad & [\,\mathbb{K}\,,\,H\,]=-\,2\,\alpha\,B\,, \\[0pt] 
[\,D\,,\,B\,]=(b-a)\,B\,, \qquad & [\,\mathbb{K}\,,\,P\,]=-\,2\,(\,\alpha\,\,t\,\p_t+\delta\,\,x\,\p_x\,)\,,\\[0pt]
[\,D\,,\,K\,]=(2b-a)\,K\,,\qquad & [\,\mathbb{K}\,,\,B\,]=-\,x^2\,\p_t\,(\,2\,\alpha-\delta\,)\,,\\[0pt]
[\,D\,,\,\mathbb{K}\,]=b\,\mathbb{K}\,. \nk
\eeak
These commutation relations, combined with the requirement that any additional generator on the right-hand side of the commutators which are not among the main set \eqref{cca1+1} must vanish, leads to the following result: the type $\mathbb{K}$ algebra and the type $K-\mathbb{K}$ algebra are not possible. However, the algebras of type $D$, type $K$, type $D-K$, and type $D-K-\mathbb{K}$  remain possible, as previously established. 
But, in addition to these cases, we observe that in $1+1$ dimensions the type $D-\mathbb{K}$ algebra can also exist. In this case, one should set $c=0$ (i.e. $K=0$) in \eqref{cca1+1}, which in turn results in choosing $\delta=2\,\alpha$ in \eqref{1+1com}. This subsequently leads to considering $a=\delta$ and $b=2\,\delta$. By following these steps, and rescaling the generators or, equivalently, setting $\delta=1$, the commutation relations \eqref{1+1com} reduce to
\beak{ll} \phantomsection \label{c-sch-a}
[\,D\,,\,H\,]=-\,H\,,\qquad &  [\,P\,,\,B\,]=H\,, \\[0pt]
[\,D\,,\,P\,]=-\,2\,P\,,\qquad &  [\,\mathbb{K}\,,\,H\,]=-\,B\,, \\[0pt] 
[\,D\,,\,B\,]=B\,,\qquad & [\,\mathbb{K}\,,\,P\,]=-\,D\,,\\[0pt]
[\,D\,,\,\mathbb{K}\,]=2\,\mathbb{K}\,. \nk
\eeak
This algebra can admit a central charge $M$ such that $[\,H\,,\,B\,]=M$, which precisely corresponds to the Carroll-Schr\"odinger algebra obtained in \cite{Najafizadeh:2024imn}. We refer to the algebra in \eqref{c-sch-a} as the Carroll-Schr\"odinger algebra without central charge, and we will denote it as $\mathfrak{carrsch}(1+1)$.

In the next section, we will discuss infinite-dimensional algebras in any dimension. However, since the Carroll-Schr\"odinger algebra exists only in two dimensions, we will present its infinite-dimensional version here to complete our discussion on this type. An infinite-dimensional generalization of the algebra \eqref{c-sch-a} takes the form \cite{Najafizadeh:2024imn}
	\be 
	\boxedB{~~~
		\begin{aligned}
		&[\,L_n\,,\,L_m\,] = (n-m)\,L_{n+m}\,, \\[5pt]
		&[\,L_n\,,\,Y_\ell\,] = \le(\,\tfrac{n}{2}-\ell\,\ri)Y_{n+\ell}\,, \\[5pt]
		&[\,Y_\ell\,,\,Y_{k}\,] = 0\,,\label{infcs}
		\end{aligned}
		~~~}
	\ee
    where $ n,m\in\mathbb{Z}$, and $\ell,k\in\mathbb{Z}+\tfrac{1}{2}$. The generators satisfying this algebra are given by
\begin{align}
		L_n&=-\,x^{n+1}\,\p_x-\,\tfrac{1}{2}\,(n+1)\,x^n\,t\,\p_t\,, \\[8pt] Y_\ell&=x^{\,\ell+\frac{1}{2}}\,\p_t\,.
\end{align}
We note that the finite generators are those with $n=0, \pm \,1$ and $\ell=\pm \,\frac{1}{2}$. To make this explicit, we can write the generators in \eqref{c-sch-a} in a suggestive form
\begin{align}
&P=\p_x=-\,L_{_{-1}}\,,	\qquad &	&H=\p_t=Y_{-\frac{1}{2}}\,, \nonumber   \\[5pt]
&D=t\,\p_t+2\,x\,\p_x=-\,2L_{_0}\,, \qquad &		&B=x\,\p_t=Y_{\frac{1}{2}}\,,  \nonumber\\[5pt]
&\mathbb{K}=x\,t\,\p_t+x^2\,\p_x=-\,L_{_1}\,,\qquad &	 \label{generatorsi}
\end{align}
demonstrating that the finite-dimensional algebra \eqref{c-sch-a} is a sub-algebra of \eqref{infcs}.

\section{Infinite-dimensional conformal extensions of the Carroll algebra}\label{infdim}

The conformal Carroll algebra of level $\ell$ that was defined in \cite{Duval:2014lpa}, and denoted $\mathfrak{ccarr}_{\ell}(d+1)$, is an example of infinite-dimensional conformal extension of the Carroll algebra $\mathfrak{carr}(d+1)$.
Let us note that the dynamical exponent is twice the inverse of the so-called level, i.e. $z=2/\ell$. Since we label our algebras with $z$ rather than $\ell$, we will denote the corresponding conformal Carroll algebra after analytically continuing to non integer $\ell$ as $\infcar_{z}(d+1):=\mathfrak{ccarr}_{2/z}(d+1)$ for notational consistency.
For any $z\in\mathbb R$, this algebra has the structure of a semi-direct sum 
\begin{align}\boxed{
\infcar_{z}(d+1)=\mathfrak{conf}(S^d)\inplus^{}_{z} C^\infty(S^d)}
\end{align} 
where $\mathfrak{conf}(S^d)$ is the Lie algebra of infinitesimal conformal transformations of the celestial sphere $S^d$. The latter is realised in practice here as the space $\mathbb{R}^d$ to which one should add a point at infinity if one wants to have globally defined conformal transformations. The semi-direct sum symbol indicates that the conformal algebra $\mathfrak{conf}(S^d)$ acts on supertranslations in $C^\infty(S^d)$ while the index $z$ indicates that supertranslations transform as conformal primary scalars of scaling dimension $z$.

As will be explained in this section, we have the following hierarchy of embeddings 
\begin{equation}
\mathfrak{carr}(d+1)\subset\sclcar_z(d+1)\subset\ccar_z(d+1)\subset\infcar_z(d+1)\,,
\end{equation}
that extends \eqref{hierarchy}.
Let us stress that the conformal Carroll algebras of arbitrary level were not properly discussed in lower spacetime dimensions in \cite{Duval:2014lpa} (in the sense that the enhancement of the conformal algebra in one and two dimensions, i.e. $\mathfrak{conf}(S^1)\cong\mathfrak{diff}(S^1)$
and $\mathfrak{conf}(S^2)\cong\mathfrak{diff}(S^1)\oplus\mathfrak{diff}(S^1)$, 
was not taken into account) since the conformal Carroll algebras  $\infcar_{z}(d+1)=\mathfrak{ccarr}_{2/z}(d+1)$ were stated to always be isomorphic to $\mathfrak{so}(d+1,1)\inplus_z C^\infty(S^d)$ while this is only true for $d\geqslant 3$ if one applies the geometric definitions in \cite{Duval:2014lpa}.
In this sense, the explicit description of these algebras in two and three spacetime dimensions which will be provided below appears to be new for $z\neq 1$, to the best of the authors' knowledge.


\subsection{Spatial conformal Carroll  algebra (\texorpdfstring{$z=0$}{})}

As explained in Section \ref{tyK}, we found the finite-dimensional conformal Carroll algebra of type K with commutation relations \eqref{tK}. Here, we investigate the possibility of extending this algebra into an infinite-dimensional algebra within the context of two, three, and higher-dimensional spacetimes.

\vspace{.5cm}
\noindent $\bullet$~\textbf{Two spacetime dimensions:} 

\noindent In $1+1$ spacetime dimensions, the rotation generator $J_{ij}$ is identically zero, and the remaining set $\{H, P,B,K\}$ of generators in \eqref{tsct}  can be denoted as the following generators
\beak{ll}\phantomsection\label{TypKinft}
H=\p_t=M_0~, \quad \qquad &  P=\p_x=-\,L_{-1}~,\\[1pt]
B=x\,\p_t=M_1~, \quad \qquad & K=x^2\,\p_t= M_2~.\nk 
\eeak
The finite-dimensional algebra of type K in \eqref{tK}, which simplifies to the commutation relations $[\,P\,,\,B\,]=H$ and $[\,P\,,\,K\,]=2\,B$, upon the identification \eqref{TypKinft}, can be equivalently re-expressed as
\be 
[\,L_{-1}\,,\,M_1\,]=-\,M_0\,,\qquad\qquad[\,L_{-1}\,,\,M_2\,]=-\,2\,M_1\,. \label{2dtk}
\ee
One can also add the dilatation operator at $z=0$ which has similar relations in $d=1$ from \eqref{tDK}. The generators of this algebra, $M_{0,1,2}$ and $L_{-1,0}$, are labeled by integer values. As a result, the algebra \eqref{2dtk} can conveniently be extended to an infinite-dimensional Lie algebra with commutation relations
\be
\boxed{~~~
\begin{aligned}
&[\,L_n\,,\,L_m\,] = (n-m)\,L_{n+m}\,, \\
&[\,L_n\,,\,M_m\,] = -\,m\,M_{n+m}\,, \qquad \qquad\qquad n,m\in\mathbb{Z}\,, \\
&[\,M_n\,,\,M_m\,] = 0\,, \label{2ditk}
\end{aligned}
~~~}
\ee
in which the infinite-dimensional generators are
\be
\boxed{~~~L_n=-\,x^{n+1}\,\p_x\,, \qquad \qquad M_n=x^n\,\p_t\,,~~~} \label{idgenera}
\ee 
which correspond to infinitesimal diffeomorphisms of the circle and to infinitesimal supertranslations, respectively. The algebra \eqref{2ditk} is the semi-direct sum of a Witt algebra and a $U(1)$ current algebra. The
central extension of this symmetry algebra is denoted as warped conformal symmetry which
contains a Virasoro algebra and a $U(1)$ Kac-Moody algebra \cite{Hofman:2011zj,Detournay:2012pc,Afshar:2015wjm}.
Note that among these generators one may identify the dilatation generator with dynamical exponent $z=0$ as follows: $D=x\,\partial_x=-L_0$. In this sense, the above algebra is necessarily an infinite-dimensional extension of the finite-dimensional conformal Carroll algebra $\ccar_0(1+1)$.

As a side remark, let us point out that one may also identify among the generators the spatial SCT generator, $\mathbb{K}=x^2\partial_x=-L_1$. However, the commutation relation $[\,L_1\,,\,M_m\,] = -\,m\,M_{m+1}$ shows that there is no finite-dimensional truncation of the algebra that would include both the spatial SCT generator and the boost generator $B=x\,\p_t=M_1$. This confirms the observation in Section \ref{tyDKKi}. 

Inspired by the terminology of \cite{Duval:2014lpa}, the algebra spanned by the generators $\{L_m,M_n\}$ will be called the conformal Carroll algebra of vanishing dynamical exponent in two dimensions, denoted $\infcar_0(1+1)$. It is a semi-direct sum $\mathfrak{diff}(S^1)\inplus C^\infty(S^1)$ where, strictly speaking, we allow vector fields and functions on the circle which are Laurent series (rather than smooth as the notation suggests). Since $z=0$, in this semi-direct sum the spatial vector fields act on spatial functions in the natural way.

\vspace{.5cm}
\noindent $\bullet$~\textbf{Three spacetime dimensions:} 

\noindent In $2+1$ spacetime dimensions, the generators of the conformal Carroll algebra of type K \eqref{tsct} are given by
\begin{equation}\label{2gentk}
H=\p_t\,,\qquad P_i=\p_i \,, \qquad B_i=x_i\,\p_t\,,\qquad  J =\epsilon^{ij} x_i\,\p_j\,, \qquad K=x^ix_i\,\p_t\,,    
\end{equation} 
where $i=1,2$. Therefore, the finite-dimensional algebra \eqref{tK} simplifies to 
\beak{ll}\phantomsection\label{3rd commutator}
[\,P_i\,,\,B_j\,]=\d_{ij}\,H\,,\quad \qquad & [\,P_i\,,\,J\,]=\epsilon_{ij}\,P^j\,,\\[5pt]
[\,K\,,\,P_i\,]=-\,2\,B_i\,,\quad \qquad & [\,B_i\,,\,J\,]=\epsilon_{ij}\,B^j\,,\nk   
\eeak
where $\epsilon_{ij}$ is the Levi-Civita symbol. In order to arrive at an infinite-dimensional algebra, let us change the basis. Accordingly, we express the spatial directions, $x_1$ and $x_2$, in terms of complex coordinates $w=x_1+i\,x_2$ and $\bar{w}=x_1-i\,x_2$. This in turn demonstrates 
\beak{lll}\phantomsection
 x_1=\frac{1}{2}\,(w+\bar{w}) \,,\qquad\qquad & \p_1=\p_w+\p_{\bar{w}} \,,\qquad\qquad & \p_w=\frac{1}{2}\,(\p_1-i\,\p_2)\,, \\[5pt]
 x_2=\frac{1}{2i}\,(w-\bar{w})	\,,\qquad\qquad & \p_2=i(\p_w-\p_{\bar{w}})\,,\qquad\qquad & \p_{\bar{w}}=\frac{1}{2}\,(\p_1+i\,\p_2)\,. \label{2gentk3}\nk
\eeak
Using the complex coordinates $w$, $\bar{w}$, we can rewrite the generators \eqref{2gentk} in terms of these variables
\beak{ll}\phantomsection
H=\p_t=M_{(0,0)} \,,\qquad\qquad &  P_1=\p_w+\p_{\bar{w}}=-\,(L_{-1}+\bar{L}_{-1})\,,\\[8pt]
B_1=\tfrac{1}{2}\,(w+\bar{w})\,\p_t=\tfrac{1}{2}\,\Big[M_{(1,0)}+M_{(0,1)}\Big]\,,\qquad\qquad & P_2=i(\p_w-\p_{\bar{w}})=-\,i(L_{-1}-\bar{L}_{-1})  \,,\\[8pt]
 B_2=\tfrac{1}{2i}\,(w-\bar{w})\,\p_t=\tfrac{1}{2i}\,\Big[M_{(1,0)}-M_{(0,1)}\Big] \,,\qquad\qquad & J=i(w\p_w-\bar{w}\p_{\bar{w}})=-\,i\,(L_0-\bar{L}_0) \,,\\[8pt]
K= w\,\bar{w}\,\p_t=M_{(1,1)} \,,\qquad\qquad &   \label{gene1} \nk
\eeak
where the second equalities, for the expression of each generator, provide the identification we applied. As a result of this, for example, the third commutator in \eqref{3rd commutator}, for $i=1$, becomes    
\be 
[\,K\,,\,P_1\,]=-\,2\,B_1 ~~~~~~\longrightarrow~~~~~~ [\,M_{(1,1)}\,,\,L_{-1}+\bar{L}_{-1}\,]=M_{(1,0)}+M_{(0,1)}\,.
\ee  
Therefore, by identifying the generators as expressed in \eqref{gene1}, we can extend these generators to the infinite-dimensional version
\be
\boxed{~~~
	\begin{aligned}
	& M_{(r,s)}:=w^{r}\,\bar{w}^{s}\,\p_t \,,\\[9pt]
	& L_n:= -\, w^{\,n+1}\,\p_w \,, \\[9pt]
	& \bar{L}_n:= -\,\bar{w}^{\,n+1}\,\p_{\bar{w}} \,,
\end{aligned}
	~~~}
\ee
where $m,n,r,s \in\mathbb{Z}$. They respectively correspond to infinitesimal supertranslations and to infinitesimal conformal transformations of the sphere. In particular, one may identify among the tower of generators the dilatation and spatial SCT generators as follows
\begin{align}
 K_1&=w^2\p_w+\bar{w}^2\p_{\bar{w}}=-\,(L_1+\bar{L}_1)\,,\qquad\quad D=-L_0=-\bar{L}_0\,,\nn\\[5pt]
 K_2&=i(w^2\p_w-\bar{w}^2\p_{\bar{w}})=-\,i(L_1-\bar{L}_1) 
\end{align}
One can truncate the infinite set of generators to the finite set of generators of the type D-K algebra. In this sense, the above algebra is indeed an infinite-dimensional extension of the finite-dimensional conformal Carroll algebra $\ccar_0(2+1)$ in three spacetime dimensions. However, the commutation relations $[\,L_1\,,\,M_{(r,s)}\,] =-\,r\,M_{(r+1,s)}$ and $[\,\bar{L}_1\,,\,M_{(r,s)}\,] =-\,s\,M_{(r,s+1)}$ show that the inclusion of spatial SCT generators $K_i$ implies that one will generate all supertranslations $M_{(r,s)}$ by taking commutators with the boosts $B_j$ and their descendants.

The whole set of generators at $z=0$ satisfies the following commutation relations 
\be
\boxed{~~~
	\begin{aligned}
		&[\,L_n\,,\,L_m\,] = (n-m)\,L_{n+m}\,,  \\[9pt]
		&[\,\bar{L}_n\,,\,\bar{L}_m\,] = (n-m)\,\bar{L}_{n+m}\,,\\[9pt]
		&[\,L_n\,,\,M_{(r,s)}\,] =-\,r\,M_{(r+n,s)} \,, \, \qquad \qquad~~~~~m,n, r,s \in\mathbb{Z}\,, \\[9pt]
            &[\,\bar{L}_n\,,\,M_{(r,s)}\,] =-\,s\,M_{(r,s+n)} \,, \, \qquad \qquad\\[9pt]
		&[\,M_{(r,s)}\,,\,M_{(t,u)}\,] = 0\,. \label{3dtkcca}
	\end{aligned}
	~~~}
\ee
Following the geometrical definition of \cite{Duval:2014lpa}, this algebra identifies with the conformal Carroll algebra of vanishing dynamical exponent in three dimensions, i.e. $\infcar_0(2+1)=\mathfrak{ccarr}_\infty(2+1)$. It is a semi-direct sum $\mathfrak{conf}(S^2)\inplus C^\infty(S^2)$ with $\mathfrak{conf}(S^2)\cong\mathfrak{diff}(S^1)\oplus\mathfrak{diff}(S^1)$, where again, strictly speaking, we consider vector fields and functions on the sphere which are Laurent series (rather than smooth as the notation suggests). This comment will apply to the subsequent subsections.
This algebra is a spatial scaling ($z=0$) analogue of the extended BMS algebra $\mathfrak{ebms}_4$ \cite{Barnich:2009se,Barnich:2010eb,Barnich:2010ojg} since it includes super-rotations.

\vspace{.5cm}
\noindent $\bullet$~\textbf{Higher spacetime dimensions ($d\geqslant 3$):} 

\noindent Let us divide the generators of the conformal Carroll algebra of type K \eqref{tsct} into two sets of generators $\{P_i, J_{ij}\}$ and $\{H, B_i, K\}$. The former set does not possess an infinite-dimensional extension, while the latter set can be infinitely extended by supertranslation generators \cite{Basu:2018dub}
\be 
M_f=f(x_i)\,\partial_t\,, \label{st} 
\ee 
where $f(x_i)$ are arbitrary tensors expressed as functions of the original spatial coordinates $x_i$ ($i=1,\ldots,d$). We noted earlier that the choices of $f$ being 1, $x_i$ or $x_j x^j$ result in $M_1=H$, $M_{x_i}=B_i$ or $M_{x^2}=K$, respectively. Therefore, for higher spacetime dimensions ($d \geqslant 3$), the infinite-dimensional extension of the type K algebra encompasses the following non-zero commutation relations 
\be
\boxed{~~~
	\begin{aligned}
		&[\,J_{ij}\,,\,M_f\,] = M_g\,,  \qquad\qquad g=J_{ij}\,f\,,\\[8pt]
		&[\,P_i\,,\,M_f\,] =M_{g^\prime}\,,\qquad\qquad g^\prime=P_i\,f\,. \label{i>4tk}\nk
	\end{aligned}
	~~~}
\ee
The bracket of the generators within the set $\{P_i, J_{ij}\}$ will remain as the ones in \eqref{tK}. 
This conformal Carroll algebra of vanishing dynamical exponent is the semi-direct sum 
\begin{align}\boxed{
\infcar_0(d+1)=\mathfrak{iso}(d+1)\inplus C^\infty(\mathbb{R}^d)\,,\qquad d \geqslant 3}
\end{align}
where the isometries of the Euclidean space act on the functions thereon and corresponds exactly to the conformal Carroll algebra of infinite level $\mathfrak{ccarr}_\infty(d+1)$, as described in \cite{Duval:2014lpa}.

 \subsection{Conformal Carroll algebras (\texorpdfstring{$0<|z|<\infty$}{})}\label{infzarbitraryDK}

The findings of the previous section can be generalised for any dynamical exponent $0< |z|<\infty$. The case $z=\infty$ is qualitatively different and will be the subject of the next subsection.

\vspace{.5cm}
\noindent $\bullet$~\textbf{Two spacetime dimensions:} 

\noindent In $1+1$ spacetime dimensions, the conformal Carroll algebra of type D-K with dynamical exponent $z$, namely $\ccar_z(1+1)$, has the infinite-dimensional extension denoted as $\infcar_z(1+1)$ with  commutation relations
\be
\boxed{~~~
\begin{aligned}
&[\,L_n\,,\,L_m\,] = (n-m)\,L_{n+m}\,, \\[5pt]
&[\,L_n\,,\,M_r\,] = \big((n+1)\,z-r\big)\,M_{n+r}\,, \qquad (n,m,r,s\in\mathbb{Z})\qquad \label{confcarrz} \\[5pt]
&[\,M_r\,,\,M_s\,] = 0\,,
\end{aligned}
~~~}
\ee
where
\be
\boxed{~~~L_n=-\,z\,(n+1)\,x^n\,t\,\p_t\,-\,x^{n+1}\,\p_x\,, \qquad \qquad M_r=x^{\,r}\,\p_t\,.~~~}
\ee 
In the terminology of \cite{Duval:2014lpa}, this algebra is called the ``conformal Carroll algebra of level $2/z$ in two dimensions''.{\footnote{It is the prototype of the Lie algebras, denoted as $W(a,b)$ in the literature, and defined as the semi-direct sum of the Witt algebra with supertranslation generators $M_n$ having non-zero commutator $[L_n,M_m]=-(a+m+bn)M_{m+n}$ (where, $a=b=-z$ here). For deformation or extension of $W(a,b)$-algebras, see \cite{FarahmandParsa:2018ojt,Safari:2020pje,Enriquez-Rojo:2021hna}.}${}^,$\footnote{Famous examples of such algebras are $\mathfrak{bms}_3$ at $z=1$, warped-witt at $z=0$ and $\mathfrak{bms}_2$ at $z=-1$ \cite{Afshar:2021qvi}. The notation $\mathfrak{bms}_2$ is used since it is the asymptotic symmetry of 2D flat-space JT gravity in analogy with its 3D counterpart $\mathfrak{bms}_3$. The algebra \eqref{confcarrz} at $z=\pm1$ coincides with the $\mathfrak{bms}_3$ and $\mathfrak{bms}_2$ algebras after shifting $r\to r\pm1$ and renaming $M_{r}\to M_{r\mp1}$ respectively.\label{bmsfootnote}}}

Since $J_{ij}=0$, the set $\{H, B, P, D,K\}$ of generators of the type D-K algebra $\ccar_z(1+1)$  can be identified as the following generators
\beak{ll}\phantomsection
H=\p_t=M_0~, \quad \qquad &   P=\p_x=-\,L_{-1}~,\\[5pt]
B=x\,\p_t=M_1~, \quad\qquad & D=z\,t\,\p_t+\,x\,\p_x=-\,L_0~, \label{gentd}\\[5pt]
K=x^2\p_t=M_2~.& \nk
\eeak

The spatial special conformal transformation generator can be identified as the generator $\mathbb{K}=2z \,x\,t\p_t+x^2\p_x=-L_1$.
The commutation relation $[\,L_1\,,\,M_r\,] = \big(2z-r\big)\,M_{r+1}$ shows that, for any integer or half-integer dynamical exponent, i.e. $z=\frac{N}2$ with $N\in\mathbb N$, one can both include the spatial SCT generator and consistently truncate the supertranslation generators $M_n$ to the finite collection with $0\leqslant n\leqslant N$ since $[\,L_1\,,\,M_N\,] = 0$. The Lie algebra spanned by the $\{L_{-1},L_0,L_1,M_0,M_1,\cdots, M_N\}$ will be denoted $\mathfrak{\cca}_{N/2}(1+1)$ and called the extended Carrollian conformal algebra with dynamical exponent $z=\frac{N}2$. It has the structure of a semi-direct sum $\mathfrak{so}(2,1)\inplus \mathbb{R}^{N+1}$.  Note that the case $N=1$ in two dimensions is special in the sense that it is the only truncation which is closed without temporal SCT $K$ and all $N>1$ truncated algebras include both temporal and spatial SCT's.
For $N=1$ ($z=1/2$), this gives the Carroll-Schr\"{o}dinger algebra found in \cite{Najafizadeh:2024imn} with supertranslation generators $M_0, M_1$ corresponding to $Y_{\pm1/2}$  as in \eqref{infcs}.  For $N=2$ ($z=1$), this gives the Carrollian conformal algebra $\cca_1(1+1)$  with supertranslation generators $M_0, M_1, M_2$ corresponding to $M_{\pm1,0}$ in the $\mathfrak{bms}_3$ algebra, see  footnote \ref{bmsfootnote}.

\vspace{.5cm}
\noindent $\bullet$~\textbf{Three spacetime dimensions:}

\noindent In $2+1$ spacetime dimensions, the generators of the conformal Carroll algebra of type D-K are given by
\beak{lll}
H=\p_t\,,\qquad\qquad & P_i=\p_i \,, \qquad\qquad &  B_i=x_i\,\p_t\,,\qquad\qquad  J =\epsilon^{ij}\,x_i\,\p_j\,, \\[5pt]
D=z\,t\p_t+x^i\p_i\,, \qquad\qquad & K=x^2\p_t\,, \nk
\eeak
where $i=1,2$. Then, the algebra becomes
\beak{lll}\phantomsection
[\,P_i\,,\,B_j\,]=\d_{ij}\,H\,,\qquad &   [\,D\,,\,H\,]=-\,z\,H\,,\qquad &[\,D\,,\,K\,]=(2-z)\,K\,,\\[3pt]
[\,P_i\,,\,J\,]=\epsilon_{ij}\,P^j\,,\qquad & [\,D\,,\,P_i\,]=-\,P_i\,,\qquad & [\,K\,,\,P_i\,]=-\,2\,B_i\,,\\[3pt]
[\,B_i\,,\,J\,]=\epsilon_{ij}\,B^j\,,\qquad &  [\,D\,,\,B_i\,]=(1-z)\,B_i\,. \qquad &\nk
\eeak
In component form, the generators are
\beak{ll}\phantomsection
P_1=\p_1 \,,\qquad\qquad &  H=\p_t\,,\\[3pt]
P_2=\p_2 \,,\qquad\qquad &  B_1=x_1\,\p_t \,,\\[3pt]
J=x_1\p_2-x_2\p_1 \,,\qquad\qquad &  B_2=x_2\,\p_t \,,\\[3pt]
D=z\,t\p_t+x_1\p_1+x_2\p_2 \,.\qquad\qquad & K=\big((x_1)^2+(x_2)^2)\p_t  \label{gen} \nk
\eeak
We can now write the spatial coordinates 
in terms of complex coordinates, $w=x_1+i\,x_2$ and $\bar{w}=x_1-i\,x_2$. Therefore, in terms of complex coordinates, the generators \eqref{gen} can then be identified as follows
\beak{ll}\phantomsection
P_1=\p_w+\p_{\bar{w}}=-\,(L_{-1}+\bar{L}_{-1}) \,,\qquad\qquad &  H=\p_t=M_{(0,0)}\,,\\[5pt]
P_2=i(\p_w-\p_{\bar{w}})=-\,i(L_{-1}-\bar{L}_{-1})\,,\qquad\qquad &  B_1=\frac{1}{2}\,(w+\bar{w})\,\p_t=\frac{1}{2}\,\Big[M_{(1,0)}+M_{(0,1)}\Big] \,,\\[5pt]
J=i(w\p_w-\bar{w}\p_{\bar{w}})=-\,i\,(L_0-\bar{L}_0) \,,\qquad\qquad &  B_2=\frac{1}{2i}\,(w-\bar{w})\,\p_t=\frac{1}{2i}\,\Big[M_{(1,0)}-M_{(0,1)}\Big] \,,\\[5pt]
D=z\,t\p_t+w\p_w+\bar{w}\p_{\bar{w}}=-\,(L_0+\bar{L}_0) \,,\qquad\qquad &  K=|w|^2\p_t=-M_{(1,1)} \label{gene11} \nk
\eeak
in which we introduced the notation
\be
\boxed{~~~
	\begin{aligned}
	& M_{(r,s)}:=w^{r}\,\bar{w}^{s}\,\p_t \,,\\[9pt]
	& L_n:= -\,\frac{z}2\,(n+1)\,w^{\,n}\,t\,\p_t- w^{\,n+1}\,\p_w \,, \\[9pt]
	& \bar{L}_n:= -\,\frac{z}2\,(n+1)\,\bar{w}^{\,n}\,t\,\p_t- \bar{w}^{\,n+1}\,\p_{\bar{w}} \,.
\end{aligned}
	~~~}
\ee
This infinite collections of generators satisfy the conformal Carroll algebra $\infcar_z(2+1)$ of dynamical exponent $z$ in three spacetime dimensions
\be
\boxed{~~~
	\begin{aligned}
		&[\,L_n\,,\,L_m\,] = (n-m)\,L_{n+m}\,,  ~~\, \qquad\qquad\qquad n,m\in\mathbb{Z}\,,\\[9pt]
		&[\,\bar{L}_n\,,\,\bar{L}_m\,] = (n-m)\,\bar{L}_{n+m}\,,\\[9pt]
		&[\,L_n\,,\,M_{(r,s)}\,] =\le(\,\frac{z}2(n+1)-r\ri)M_{(r+n,s)} \,, \, \qquad  \\[9pt]
            &[\,\bar{L}_n\,,\,M_{(r,s)}\,] =\le(\,\frac{z}2(n+1)-s\ri)M_{(r,s+n)} \,, \, \qquad \qquad\\[9pt]
		&[\,M_{(r,s)}\,,\,M_{(t,u)}\,] = 0\,,~~\,  \qquad\qquad\quad\qquad\qquad r,s,t,u\in\mathbb{Z}\,.\label{3dcca}
	\end{aligned}
	~~~}
\ee
Let us note once again that spatial SCT generators $K_i$ correspond to $L_1$ and $\bar{L}_1$. Their commutation relations with $M_{(r,s)}$ will produce $(z-r)M_{(r+1,s)}$ and $(z-s)M_{(r,s+1)}$, respectively, in the right-hand side. In particular, $[\,L_1\,,\,M_{(1,0)}\,] =(z-1)M_{(2,0)}$ and  $[\,\bar{L}_1\,,\,M_{(0,1)}\,] =(z-1)M_{(0,2)}$. As one can see explicitly, the inclusion of the special conformal transformation will produce new generators, except if $z=1$.

For $z=1$, the algebra \eqref{3dcca} becomes isomorphic to the extended BMS algebra, i.e. $\mathfrak{ebms}_4\cong \infcar_1(2+1)$:
\begin{align}
		&[\,L_n\,,\,L_m\,] = (n-m)\,L_{n+m}\,, ~~\, \nonumber\\[6pt]
		&[\,\bar{L}_n\,,\,\bar{L}_m\,] = (n-m)\,\bar{L}_{n+m}\,,\nonumber\\[6pt]
		&[\,L_n\,,\,M_{(r,s)}\,] =\le(\frac{n+1}{2}-r\ri)M_{(r+n,s)} \,, \, \qquad\qquad n,m,r,s,t,u\in\mathbb{Z}\,,\nonumber\\[6pt]
            &[\,\bar{L}_n\,,\,M_{(r,s)}\,] =\le(\frac{n+1}{2}-s\ri)M_{(r,s+n)} \,, \, \qquad \qquad\nonumber\\[6pt]
		&[\,M_{(r,s)}\,,\,M_{(t,u)}\,] = 0\,.\label{bms4}
\end{align}
This algebra of course admits as a finite-dimensional truncation, $\cca_1(2+1)\cong\mathfrak{iso}(3,1)$, the Carrollian conformal algebra in three dimensions spanned by the zero modes $0,\pm1$. Indeed, the finite subset $\{H,B_1,B_2,K\}$ of supertranslations spans the finite-dimensional vector space $\mathbb{R}^{3,1}$ which is the vector representation of the Lie algebra $\mathfrak{so}(3,1)$ spanned by $\{P_i,J_{ij},K_j,D\}$.

Let us observe that, more generally for any integer dynamical exponent $z=N\in\mathbb N$, there exists a finite-dimensional truncation of the conformal Carroll algebra with finitely-many supertranslation generators. Nevertheless, when $N>1$ the price to pay is that one has to include supertranslations beyond the standard ones $\{H,B_i,K\}$. In fact, note that for $z=N$ the commutation relations \eqref{3dcca} imply that
$[\,L_1\,,\,M_{(N,s)}\,] =0$ and $[\,\bar{L}_1\,,\,M_{(r,N)}\,] =0$. Therefore, one can consistently  truncate the supertranslation generators $M_{(r,s)}$ to the ones with $0\leqslant r,s\leqslant N$. We call these finite-dimensional algebras the extended Carrollian conformal algebras and denote them as $\cca_N(2+1)$. This subset spans a finite-dimensional representation of $\mathfrak{so}(3,1)$ of dimension $(N+1)^2$ which is the tensor product $D^{\mathfrak{so}(3)}_{\frac{N}2}\otimes D^{\mathfrak{so}(3)}_{\frac{N}2}$, where $D^{\mathfrak{so}(3)}_S$ denotes the representation the rotation algebra $\mathfrak{so}(3)$ of dimension $2S+1$. Equivalently, the supertranslation generators $\{M_{(r,s)}\}_{r,s\geqslant N}$ span the irreducible representation $D^{\mathfrak{so}(3,1)}_N$ of Lorentz algebra $\mathfrak{so}(3,1)$ spanned by symmetric traceless Lorentz tensors of rank $N$. As usual, this follows from the exceptional isomorphism $\mathfrak{so}_\mathbb{C}(4)\cong \mathfrak{so}_\mathbb{C}(3)\oplus\mathfrak{so}_\mathbb{C}(3)$ of complex algebras.

\vspace{.5cm}
\noindent $\bullet$~\textbf{Higher spacetime dimensions ($d\geqslant3$):} 

\noindent The finite-dimensional generators of the conformal Carroll algebras of type D-K consist of the set of generators $\{H,P_i, J_{ij},B_j, D,K\}$. The subset $\{H,B_i\,,K\}$ has an infinite-dimensional extension, which is provided by the supertranslation generators $M_f=f(x^i)\,\p_t$, as introduced previously in \eqref{st}. As a result, the infinite-dimensional extension of the type D and type D-K algebras become
\be
\boxed{~~~
	\begin{aligned}
		&[\,J_{ij}\,,\,M_f\,] = M_g\,,  \qquad\qquad g=J_{ij}\,f\,,\\[8pt]
		&[\,P_i\,,\,M_f\,] =M_{g^\prime}\,,\qquad\qquad g^\prime=P_i\,f\,,\\[8pt]
            &[\,D\,,\,M_f\,] =M_{h}\,,\qquad\qquad ~h=(x\cdot\p-z)\,f\,. \label{i>4td}\nk
	\end{aligned}
	~~~}
\ee
This reproduces the conformal Carroll algebra $\mathfrak{ccarr}_{2/z}(d+1)=\infcar_{z}(d+1)$ of level $2/z$ or, equivalently, of dynamical exponent $z$. 

When the dynamical exponent is an integer $z=N$, we can consistently  truncate the supertranslation algebra to the finite-dimensional representation $D_N$ (symmetric traceless tensors of rank $N$) of the Lorentz algebra $\mathfrak{so}(d+1,1)$ spanned by $\{P_i,J_{ij},K_j,D\}$.
The corresponding truncation is the algebra $\cca_N(d+1)=\mathfrak{so}(d+1,1)\inplus D_N$.
We have the following hierarchy of Lie subalgebras that generalises \eqref{bmshierarch} to all integers $N\geqslant 1$:
\begin{equation}
\mathfrak{carr}(d+1)\subset\sclcar_N(d+1)\subset\ccar_N(d+1)\subset\cca_N(d+1)\subset\infcar_N(d+1)\,.
\end{equation}
Remember that for $N>1$, these extended conformal Carroll algebras include supertranslation generators beyond $H$, $B_i$ and $K$.

 \subsection{Temporal conformal Carroll algebra (\texorpdfstring{$|z|=\infty$}{})}

The findings of the previous section can be adapted to the limit $z\to\pm\infty$ as follows.

\vspace{.5cm}
\noindent $\bullet$~\textbf{Two spacetime dimensions:} 

\noindent In $1+1$ spacetime dimensions, consider the infinite-dimensional algebra with commutation relations 
\be
\boxed{~~~
\begin{aligned}
&[\,L_n\,,\,L_m\,] = 0\,, \\[5pt]
&[\,L_n\,,\,M_r\,] = M_{n+r}\,, \qquad (n,m,r,s\in\mathbb{Z})\qquad  \\[5pt]
&[\,M_r\,,\,M_s\,] = 0\,,
\end{aligned}
~~~}
\ee
where
\be
\boxed{~~~L_n=\,-\,x^n\,t\,\p_t\,, \qquad \qquad M_r=x^{\,r}\,\p_t\,.~~~}
\ee 
Since $J_{ij}=0$, the subset $\{H, B, D,K\}$ of generators of the temporal conformal Carroll algebra $\ccar_\infty(1+1)$  can be identified as the following generators
\beak{ll}\phantomsection
H=\p_t=M_0~, \quad \qquad & K=x^2\p_t=M_2 ~,\\[5pt]
B=x\,\p_t=M_1~, \quad\qquad & D=\,t\,\p_t=-\,L_0~, \label{gentd2}\nk
\eeak
As one can see, the spatial translation generator $P=\partial_x$ must be added by hand in order for the infinite-dimensional algebra to be an extension of the Carroll algebra $\mathfrak{carr}(1+1)$. This only produces two extra nontrivial commutation relations
\be
\boxed{~~~
\begin{aligned}
&[\,P\,,\,L_n\,] = n\,L_{n-1}\,, \\[5pt]
&[\,P\,,\,M_r\,] = r\,M_{r-1}\,. \qquad (n,r\in\mathbb{Z})\qquad  \\[5pt]
\end{aligned}
~~~}
\ee

\vspace{.5cm}
\noindent $\bullet$~\textbf{Three spacetime dimensions:} 

\noindent In $2+1$ spacetime dimensions, the generators of the infinite-dimension extension of the temporal conformal Carroll algebra $\ccar_\infty(2+1)$ are taken to
\be
\boxed{~~~
	\begin{aligned}
 P_1&=\p_w+\p_{\bar{w}}\,,\quad\qquad\qquad\; L_n:= -\,w^{\,n}\,t\,\p_t\,,\\[9pt]
P_2&=i(\p_w-\p_{\bar{w}})\,,\quad\qquad\quad\;\bar{L}_n:= -\,\bar{w}^{\,n}\,t\,\p_t\,,\\[9pt]
J&=i(w\p_w-\bar{w}\p_{\bar{w}}) \,,\quad\;\,\, M_{(r,s)}:=-\,w^{r}\,\bar{w}^{s}\,\p_t \,.
\end{aligned}
	~~~}
\ee
whose commutation relations read
\be
\boxed{~~~
	\begin{aligned}
		&[\,L_n\,,\,L_m\,] = 0= [\,\bar{L}_n\,,\,\bar{L}_m\,]\,,\quad\, [\,J\,,\,L_n\,]=inL_n\,, \qquad[\,P_k\,,\,L_n\,]= i^{k+1}nL_{n-1}\,, \\[9pt]
		&[\,L_n\,,\,M_{(r,s)}\,] =M_{(r+n,s)} \,, \, \qquad \;\, [\,J\,,\,\bar L_n\,]=-in\bar L_n\,,\quad\,[\,P_k\,,\,\bar L_n\,]=i^{k+1}n\bar L_{n-1}\,,\\[9pt]
            &[\,\bar{L}_n\,,\,M_{(r,s)}\,] =M_{(r,s+n)} \,, \; \quad [\,J\,,\,M_{(r,s)}\,]=i(r-s)M_{(r,s)}\,,\quad[\,J\,,\,P_k\,]=\epsilon_{k\ell}P_\ell\,,\\[9pt]
		&[\,M_{(r,s)}\,,\,M_{(t,u)}\,] = 0\,, \qquad\;\;[\,P_k\,,\,M_{(r,s)}\,]=i^{k+1}(rM_{(r-1,s)}+(-1)^{k-1}sM_{(r,s-1)}) \,.
	\end{aligned}
	~~~}
\ee
where $n,m,r,s,t,u\in\mathbb{Z}$ and $k,\ell=1,2$ with $\epsilon_{21}=1$.
The subset $\{H,B_i,D\}$ of generators of the temporal conformal Carroll algebra $\ccar_\infty(2+1)$ can then be identified as
\beak{ll}\phantomsection
B_1=\frac{1}{2}\,(w+\bar{w})\,\p_t=\frac{1}{2}\,\Big[M_{(1,0)}+M_{(0,1)}\Big] ~, \quad \qquad & D=\,t\,\p_t=-\,L_0=-\,\bar{L}_0~,\\[8pt]
 B_2=\frac{1}{2i}\,(w-\bar{w})\,\p_t=\frac{1}{2i}\,\Big[M_{(1,0)}-M_{(0,1)}\Big]~. &  \nk
\eeak

\vspace{.5cm}
\noindent $\bullet$~\textbf{Higher spacetime dimensions ($d\geqslant3$):} 

\noindent In any dimension, the basic idea is to extend the generators $H$ and $B_i$ to the supertranslations $M_f=f(x^i)\,\p_t$ and the dilatation generator $D=t\p_t$. Together with the set $\{P_i,J_{ij}\}$ of generators of the Euclidean algebra $\mathfrak{iso}(d)$, this gives an infinite-dimensional extension $\Widetilde{\mathfrak{confcarr}}_\infty(d+1)$ of the temporal conformal Carroll algebra $\mathfrak{confcarr}_\infty(d+1)$. This algebra $\Widetilde{\mathfrak{confcarr}}_\infty(d+1)$  is a Lie subalgebra of the conformal algebra of vanishing level $\mathfrak{ccarr}_0(d+1)$ discussed in \cite{Duval:2014lpa}. In the latter algebra, the supertranslation and dilatation generators are generalised even further to $M_f=f(x^i,t)\partial_t$.

The infinite-dimensional extensions of the type D-K-K$_i$ algebra, i.e. the Carrollian conformal algebra, have been extensively studied in the literature. See, in particular, \cite{Duval:2014lpa} for its BMS extension in any dimension and \cite{Barnich:2009se,Barnich:2010eb,Barnich:2010ojg}  for its extension by super-rotations when $d\leqslant 3$ (see also \cite{Bagchi:2019xfx,Bagchi:2016bcd}). This type of extension is included already in our generic discussion of section \ref{infzarbitraryDK} after setting $z=1$. 

\section{Ward identities and correlation functions} \label{wardidentity}
 
Let us consider $n$-point correlation functions of primary fields (with respect to the conformal Carroll algebra under consideration).
We want to investigate the Ward identities of the Carrollian CFT. They may be written as follows 
\begin{align}\label{Wardid}
	\sum_i\langle 0|\phi(x_1)\cdots [Q,\phi(x_i)]\cdots\phi(x_n)|0\rangle=0
\end{align}
where $|0\rangle$ is a vacuum which is invariant under the global part of the algebra $Q|0\rangle=0$, and $Q$ is the generator of the infinitesimal symmetry transformation via the representation of an antihermitian operator $Q$ on the field $\phi(x_i)$ such that $[Q,\phi(x_i)]=\delta\phi(x_i)$ We identify the operator $Q$ with some generator of the conformal Carroll algebras that have been investigated in the previous sections. From the antihermiticity of the operator and the invariance of the vacuum  $Q|0\rangle=0$, we can derive  eq. \eqref{Wardid}:
\begin{align}
	0&=\langle0|Q\phi(x_1)\cdots\phi(x_n)|0\rangle\nn\\
	&=\sum_i\langle0|\phi(x_1)\cdots\phi(x_{i-1})[Q,\phi(x_i)]\phi(x_{i+1})\cdots\phi(x_n)|0\rangle\nn\\
	&+\langle0|\phi(x_1)\cdots\phi(x_n)Q|0\rangle\,,
\end{align}
where the last term vanishes, and the commutator with $Q$ generates the infinitesimal transformation $\delta\phi$, hence we get \eqref{Wardid} or, equivalently,
\begin{align}\label{Qinv}
\sum_i\langle0|\phi(x_1)\cdots\phi(x_{i-1})\delta\phi(x_i)\phi(x_{i+1})\cdots\phi(x_n)|0\rangle=0\,.
\end{align}

\subsection{Two-point functions}

If we consider the two-point function 
\begin{align}
G^{(2)}(\vec x_1,t_1;\vec x_2,t_2)=\langle0|\phi_1(\vec x_1,t_1)\,\phi_2(\vec x_2,t_2)|0\rangle
\end{align}
and take the operator $Q$ in \eqref{Qinv} to be either the time translation generator $H$ or the space translation generator $\vec P$, then \eqref{Wardid} gives the respective differential equations,
\begin{align}
\left(\partial_{t_1}+\partial_{t_2}\right)G^{(2)}(\vec x_1,t_1;\vec x_2,t_2)=0\,,\qquad
(\vec\partial_{x_1}+\vec\partial_{x_2})\,G^{(2)}(\vec x_1,t_1;\vec x_2,t_2)&=0\,.
\end{align}
We thus get $G^{(2)}=G^{(2)}(\vec x_{12}, t_{12})$ where $\vec x_{12}=\vec x_1-\vec x_2$ and $t_{12}=t_1-t_2$.
If we take $Q$ as the generator of the Carrollian boost $\vec B=\vec x\,\partial_t$, then we have
\begin{align}\label{Boostdiff}
(\vec x_1\partial_{t_1}+\vec x_2\partial_{t_2})\,G^{(2)}(\vec x_{12}, t_{12})=0\,.
\end{align}
Since $\partial_{t_1}F(t_{12})=\partial_{t_{12}}F( t_{12})$ and $\partial_{t_2}F(t_{12})=-\,\partial_{t_{12}}F(t_{12})$ for any function $F$,
the differential equation \eqref{Boostdiff} is equivalent to
\begin{align}
	\vec x_{12}\,\partial_{t_{12}}\,G^{(2)}(\vec x_{12}, t_{12})=0\,,
\end{align}
which is solved as,
\begin{align}\label{Carrolian2pointpre}
	G^{(2)}(\vec x_{12}, t_{12})=G(\vec x_{12})+F(t_{12})\,\delta
 (\vec x_{12})\,,
\end{align} 
where $\delta
(\vec x):=\prod_{i=1}^d\delta(x^i)$.  Remember that the Dirac delta function has the scaling dimension of the inverse of a $d$-volume element (which is obvious since it should give a number via integration). Therefore, 
\begin{equation}\label{Diracident}
(\vec x\cdot\vec \partial_x+d\,)\,\delta
(\vec x)=0    \,.
\end{equation}
This property will be very useful below. 

The expression \eqref{Carrolian2pointpre} is invariant under the spatial rotation generator $J_{ij}$ iff it only depends on the norm $|\vec x_{12}|$, i.e. $G(\vec x_{12})=G\big(|\vec x_{12}|\big)$. In fact, the Dirac delta function is invariant under any rotation: for any rotation $R\in SO(d)$, one has $\delta(R\vec x)=\delta(\vec x)/|\det R|=\delta(\vec x)$ (since rotations have unit determinant, $\det R=1$). In this way, we get the general expression for the two-point function in a Carrollian field theory
\begin{align}\label{Carrolian2point}
	\boxed{G^{(2)}(\vec x_{12}, t_{12})=G\big(|\vec x_{12}|\big)+F(t_{12})\,\delta
 (\vec x_{12})}\,,
\end{align} 
which matches the result in \cite{deBoer:2021jej}. The first term is identical to the one of an Euclidean quantum field theory (QFT) in $d$ dimensions and is sometimes called ``magnetic''. The second term on the right-hand side \eqref{Carrolian2point} is often called ``ultra-local'' or ``electric''. It is specific to Carrollian QFT's.

\paragraph{Remark:} 
In $d=1$, the Carroll algebra admits a central term in the commutator of the generators of time translations and Carrollian boosts: $[H,B]=M$. In this case, the generator of the boost is $B=x\partial_t+Mt$ (see section \ref{csa}) and thus the representation of the algebra is projective and complex. Defining the two point function in this case as  $G^{(2)}=\langle0|\phi_1(\vec x_1,t_1)\,\phi_2^\dagger(\vec x_2,t_2)|0\rangle$, its invariance under boost, upon assuming $\delta_{_M}\phi_{1,2}=im\,\phi_{1,2}$, leads to the equation $( \vec x_{12}\,\partial_{t_{12}}+im \,t_{12}) G^{(2)}(\vec x_{12},t_{12})=0$  which is solved as $G^{(2)}(\vec x_{12},t_{12})=G(\vec x_{12}) \,\exp(-\,\frac{i m \,t_{12}^2}{2\,\vec x_{12}})+C\,\delta(t_{12})\,\delta(\vec x_{12})$, see also \cite{Najafizadeh:2024imn}. One can fix the function $G$ and $C$ upon applying the rest of conformal symmetries.

\subsubsection{K-invariance: supertranslation enhancement}

Note that the invariance under spacetime translations and Carrollian boosts implied the general form \eqref{Carrolian2pointpre} of the two-point function. Remarkably, this is enough to guarantee its invariance under generic ``supertranslations'' $t'=t+f(\vec x)$. This is trivial for the first term $G(\vec x_{12})$ since it does not depend on time. For the second term, this is because $t'_{12}=t_{12}+f(\vec x_1)-f(\vec x_2)$
implies $F(t'_{12})\,\delta(\vec x_{12})=F(t_{12})\,\delta(\vec x_{12})$ since 
\begin{equation}\label{identitydelta}
g(\vec x_1)\,\delta(\vec x_{12})=g(\vec x_2)\,\delta(\vec x_{12})
\end{equation}
for any function $g$. Therefore, free Carrollian field theories, i.e. free QFTs invariant under the Carroll group $\mathfrak{carr}(d+1)$, are automatically invariant under the infinite-dimensional Lie algebra $\mathfrak{isocarr}(d+1)$ of all Carrollian isometries.

In particular, the invariance under the Carrollian special conformal transformations generated by $K=\vec x\cdot\vec x\,\partial_t$ is equivalent to
\begin{align}\label{SCTdiff}
	0&=(\vec x_1\cdot\vec x_1\,\partial_{t_1}+\vec x_2\cdot \vec x_2\,\partial_{t_2})\left[G\big(|\vec x_{12}|\big)+F(t_{12})\delta(\vec x_{12})\right]\nn\\
	&=(\vec x_1\cdot\vec x_1-\vec x_2\cdot\vec x_2)\,F'(t_{12})\delta(\vec x_{12})
\end{align} 
which is automatically satisfied due to \eqref{identitydelta}.
Therefore, the symmetries of free Carrollian QFT's are automatically enhanced from the Carroll algebra to the type K conformal algebra $\mathfrak{Kcarr}(d+1)$. 

\subsubsection{Dilatation invariance}\label{typeDK2pt}

Now we demand further invariance, under the dilatation generator $D=z\, t\partial_t+\,\vec x\cdot\vec\partial_x+\Delta$ (the case $z=\infty$ with $D=t\partial_t+\Delta$ can be treated separately), which leads to the following differential equation,
\begin{align}
	\le(z\,t_1\partial_{t_1}+\,\vec x_1\cdot\vec\partial_{x_1}+z\,t_2\partial_{t_2}+\,\vec x_2\cdot\vec\partial_{x_2}+\Delta_1+\Delta_2\ri)\!\Big[G(\vec x_{12})+F(t_{12})\,\delta(\vec x_{12})\Big]=0\,.
\end{align} 
Since $\vec \partial_{x_1}G(\vec x_{12})=\vec\partial_{x_{12}}G(\vec x_{12})$ and $\vec \partial_{x_2}G(\vec x_{12})=-\,\vec\partial_{x_{12}}G(\vec x_{12})$ for any function $G$, this leads to 
\begin{align}\label{DinvGF}
0&=	(z\,t_{12}\partial_{t_{12}}+\,\vec x_{12}\cdot\vec\partial_{ x_{12}}+\Delta_1+\Delta_2)\big[\,G(\vec x_{12})+F(t_{12})\delta(\vec x_{12})\,\big]\nn\\[5pt]
&=z\,t_{12}F'(t_{12})\delta(\vec x_{12})+\,\vec x_{12}\cdot\vec\partial_{ x_{12}}G(\vec x_{12})-\,d\,F(t_{12})\delta(\vec x_{12})\nn\\[5pt]
&\quad+(\Delta_1+\Delta_2)\big[\,G(\vec x_{12})+F(t_{12})\delta(\vec x_{12})\,\big]\,,
\end{align} 
where we used the property \eqref{Diracident}. Rearranging eq. \eqref{DinvGF} we have
\begin{align}\label{dilat}
	\vec x_{12}\cdot\vec\partial_{ x_{12}}G(\vec x_{12})+(\Delta_1+\Delta_2)\,G(\vec x_{12})\,=\Big[(d-\Delta_1-\Delta_2)F(t_{12})-z\,t_{12}F'(t_{12})\Big]\,\delta(\vec x_{12})\,.
\end{align}
If $\vec x_{12}\neq\vec 0$, then the right hand side is zero and the homogeneous differential equation is easily solved 
as
\begin{align}\label{Gx12}
 G(\vec x_{12})\,=\,C_1\, |\vec x_{12}|^{-(\Delta_1+\Delta_2)}
 \,,
\end{align}
where $C_1$ is an arbitrary factor. 
If we insert \eqref{Gx12} in \eqref{dilat}, then we find that the left-hand side of \eqref{dilat} vanishes everywhere except possibly at $\vec x_{12}=\vec 0$. In other words, the left-hand side could be at most a (spatial) distribution with support at the origin. But the left-hand side does not depend on $t_{12}$ while the right-hand side does, this implies that 
\begin{equation}\label{FF'}
    (d-\Delta_1-\Delta_2)F(t_{12})-z\,t_{12}F'(t_{12})=0\,.
\end{equation}
Note that the Euler operator $\vec x\cdot\vec\partial_{x}=r\partial_r$ acting on a power $|\vec x|^\alpha=r^\alpha$ (with $\alpha\in\mathbb R$) never produces a Dirac delta distribution, so there is not even room for a constant in the right-hand side of \eqref{FF'}. 
For $z\neq 0$, the first-order equation \eqref{FF'} can be solved as
\begin{align}\label{Ft12}
 F(t_{12})\,=\,C_2\, |t_{12}|^{-(\Delta_1+\Delta_2-d)/z}
 \,,
\end{align}
where $C_2$ is an arbitrary factor. 
Therefore, for $z\neq\{0,\infty\}$ the two-point function has the general form
\begin{align}\label{Carrolian2pointsD}
 \boxed{G^{(2)}(\vec x_{12}, t_{12})=C_1\, |\vec x_{12}|^{-(\Delta_1+\Delta_2)}+C_2\, |t_{12}|^{-(\Delta_1+\Delta_2-d)/z}\delta(\vec x_{12})}\,.
\end{align} 
The formula \eqref{Carrolian2pointsD} is the general expression of the two-point function for two scalar fields of scaling dimensions $\Delta_1$ and $\Delta_2$ in a Carrollian CFT which is invariant under the conformal Carroll algebra of type D-K,  $\ccar_z(d+1)$.

If $z=0$, the dilatation is purely spatial and then it is clear from \eqref{FF'} that there are two possibilities;
\begin{equation}\label{Carrolian2points'}
    G^{(2)}(\vec x_{12}, t_{12})=\begin{cases}
C_1\, |\vec x_{12}|^{-(\Delta_1+\Delta_2)}, &  \Delta_1+\Delta_2\neq d\,,\\
C_1\, |\vec x_{12}|^{-d}+F(t_{12})\delta(\vec x_{12}), & \Delta_1+\Delta_2= d\,,
\end{cases}
\end{equation}
where $F(t_{12})$ is not fixed here.
These are the two possible forms of two-point functions in a Carrollian CFT which is invariant under the spatial conformal Carroll algebra $\ccar_0(d+1)$. 

If $z=\infty$, then the invariance under $D=t\partial_t+\Delta$ reads
\begin{align}
	\Big(\,t_1\partial_{t_1}+\,t_2\partial_{t_2}+\Delta_1+\Delta_2\Big)\Big[G(\vec x_{12})+F(t_{12})\delta(\vec x_{12})\Big]=0\,.
\end{align} 
which implies for $\vec x_{12}\neq\vec 0$ that either $G(\vec x_{12})=0$ if $\Delta_1+\Delta_2\neq 0$ or $G(\vec x_{12})$ is arbitrary if $\Delta_1+\Delta_2= 0$. Moreover, $(\,t_{12}\partial_{t_{12}}+\Delta_1+\Delta_2)F(t_{12})=0$ which can be solved as $F(t_{12})\,=\,C_2\, |t_{12}|^{-(\Delta_1+\Delta_2)}$. Therefore, for $z=\infty$ there are two cases to distinguish for the two-point function;
\begin{equation}\label{Carrolian2points'''}
    G^{(2)}(\vec x_{12}, t_{12})=\begin{cases}
C_2\, |t_{12}|^{-(\Delta_1+\Delta_2)}\delta(\vec x_{12}), &  \Delta_1+\Delta_2\neq0\,,\\
G(\vec x_{12})+C_2\,\delta(\vec x_{12}), & \Delta_1+\Delta_2=0\,.
\end{cases}
\end{equation}
These are the two possibilities for two-point functions in a Carrollian CFT which is invariant under the temporal conformal Carroll algebra $\ccar_\infty(d+1)$. It is interesting that in the $z=\infty$ (temporal dilatation invariance) the two-point function is independent of the space dimension $d$.

\subsubsection{K\texorpdfstring{$_i\,$}{}-invariance: Carrollian conformal symmetries}

Now let us have a look at the Carrollian conformal algebra \eqref{cca} and check how this procedure works there when we have $K_i$ as extra generators (corresponding to special conformal transformations). In this case, the generators are given by \eqref{ccagen} which leads to the same expression for the two-point function as above after applying $H, P_i, B_j, J_{ij}, K$ where $a=b=1$ since $z=1$. Nevertheless, we will treat the case of arbitrary $z$ for later purpose. The only remaining generator is $K_i=2\,x_i\,(z\,t\p_t+x^j\p_j+\Delta)-\,x^2\p_i$ under which the invariance of \eqref{Carrolian2point} gives the following differential equation;
\begin{align}
&\Big(\vec x_1(z\,t_1\partial_{t_1}+\vec x_1\cdot\vec\partial_{x_1})-\frac12(\vec x_1\cdot \vec x_1)\,\vec\partial_{x_1}+\Delta_1\,\vec x_1\nn\\&\quad +	\vec x_2(z\,t_2\partial_{t_2}+\vec x_2\cdot\vec\partial_{x_2})-\frac12(\vec x_2\cdot \vec x_2)\,\vec\partial_{x_2}+\Delta_2\,\vec x_2\Big)\,\Big[G(\vec x_{12})+F(t_{12})\delta(\vec x_{12})\Big]=0\,.
\end{align}
Again, let us look at this equation for $\vec x_{12}\neq\vec 0$ which yields:
\begin{align}
0&=\Big(\vec x_1(\vec x_1\cdot\vec\partial_{x_1})-\frac12(\vec x_1\cdot \vec x_1)\,\vec\partial_{x_1}+\Delta_1\,\vec x_1 +	\vec x_2(\vec x_2\cdot\vec\partial_{x_2})-\frac12(\vec x_2\cdot \vec x_2)\,\vec\partial_{x_2}+\Delta_2\,\vec x_2\Big)\,G(\vec x_{12})\nn
\\&=\Big(\vec x_1(\vec x_1\cdot\vec\partial_{x_{12}})-\vec x_2(\vec x_2\cdot\vec\partial_{x_{12}})-\frac12(\vec x_1\cdot\vec x_1-\vec x_2\cdot\vec x_2)\vec\partial_{x_{12}}+\Delta_1\,\vec x_1+\Delta_2\,\vec x_2\Big)G(\vec x_{12})\nn\,.    
\end{align}
This coincides with the infinitesimal form of the invariance of \eqref{Gx12} under purely spatial SCT's (because it merely is the action of the generator $K_i$ where the action of the time derivative is trivial). Plugging in the bilocal function $G(\vec x_{12})$ in \eqref{Gx12}, 
it can be checked directly that this condition is satisfied iff either $\Delta_1=\Delta_2\equiv\Delta$ or $C_1=0$.\footnote{This is a well-known fact upon  SCT invariance (see e.g. \cite[Subsection 4.3.1]{DiFrancesco:1997nk}).} Thus, in any Carrollian CFT in $d+1$ dimensions invariant under the Carrollian conformal algebra, the ``magnetic'' part of the two-point function between two scalar fields of different scaling dimensions is zero. This statement is actually true independently of the value of $z$ (including $z=0$ and $z=\infty$). Remember that one can include $K_i$ as the symmetry for general $z$ provided extra generators are also included (cf. Section \ref{infdim}).

In the coinciding limit, $\vec x_1\to\vec x_2$\,, the problem reduces to
\begin{align}&\Big(z(\vec x_1 t_1-\vec x_2 t_2)\partial_{t_{12}}+\vec x_1(\vec x_1\cdot\vec\partial_{x_1})+\vec x_2(\vec x_2\cdot\vec\partial_{x_2})\nn\\&\quad -\frac12(\vec x_1\cdot \vec x_1-\vec x_2\cdot \vec x_2)\,\vec\partial_{x_{12}}+\Delta_1\vec x_1+\Delta_2\vec x_2\Big)\,\Big[F(t_{12})\delta(\vec x_{12})\Big]=0\label{K_ix12=0}
\end{align}
Using the property $\vec x_{12}\,\delta(\vec x_{12})=0$, one gets
\begin{align}&\Big(\frac12\,\vec{\tilde x}_{12}\, (z\,t_{12}\partial_{t_{12}}+\Delta_1+\Delta_2)+\vec x_1(\vec x_1\cdot\vec\partial_{x_1})+\vec x_2(\vec x_2\cdot\vec\partial_{x_2})\nn\\&\quad -\frac12(\vec x_1\cdot \vec x_1-\vec x_2\cdot \vec x_2)\,\vec\partial_{x_{12}}\Big)\,\Big[F(t_{12})\delta(\vec x_{12})\Big]=0
\end{align}
where we introduced the notation $\vec{\tilde x}_{12}=\vec x_1+\vec x_2$. 
Taking into account \eqref{Ft12}
for $z\notin\{0,\infty\}$,
we end up with the condition
\begin{equation}
F(t_{12})\Big[\,\frac{d}2\,\vec{\tilde x}_{12}+\vec x_1(\vec x_1\cdot\vec\partial_{x_1})+\vec x_2(\vec x_2\cdot\vec\partial_{x_2})-\frac12(\vec{\tilde x}_{12}\cdot\vec x_{12})\,\vec\partial_{x_{12}}\Big]\,\delta(\vec x_{12})=0    
\end{equation}
which is satisfied for any value of $z$. 

To prove this, one can start by simplifying this expression. In order to do this, note that the property $x^i\partial_j\delta(\vec x)\,=\,-\,\delta_j^i\delta(\vec x)$ for the Dirac distribution implies that 
\begin{equation}\label{Kiidentity1}
\Big(\vec{\tilde x}_{12}\cdot\vec x_{12}\Big)\,\vec\partial_{x_{12}}\delta(\vec x_{12})=-\vec{\tilde x}_{12}\,\delta(\vec x_{12})\,.
\end{equation}
Then one can check, step by step, that
\begin{align}\label{Kiidentity2}
&\Big(\vec x_1(\vec x_1\cdot\vec\partial_{x_1})+\vec x_2(\vec x_2\cdot\vec\partial_{x_2})\Big)\,\delta(\vec x_{12})\nn\\
&\quad=\Big(\vec x_1(\vec x_1\cdot\vec\partial_{x_{12}})-\vec x_2(\vec x_2\cdot\vec\partial_{x_{12}})\Big)\,\delta(\vec x_{12})\nn\\ 
&\quad=\frac14\Big((\vec x_{12}+\vec{\tilde x}_{12})\big[(\vec x_{12}+\vec{\tilde x}_{12})\cdot\vec\partial_{x_{12}}\big]-(\vec x_{12}-\vec{\tilde x}_{12})\big[(\vec x_{12}-\vec{\tilde x}_{12})\cdot\vec\partial_{x_{12}}\big]\Big)\,\delta(\vec x_{12})\nn\\ 
&\quad=\frac14\Big((\vec x_{12}+\vec{\tilde x}_{12})(\vec{\tilde x}_{12}\cdot\vec\partial_{x_{12}}-d)+(\vec x_{12}-\vec{\tilde x}_{12})(\vec{\tilde x}_{12}\cdot\vec\partial_{x_{12}}+d)\Big)\,\delta(\vec x_{12})\nn\\ 
&\quad=\frac12\Big(\vec x_{12}\,(\vec{\tilde x}_{12}\cdot\vec\partial_{x_{12}})-d\,\vec{\tilde x}_{12}\Big)\,\delta(\vec x_{12})\nn\\ 
&\quad=\frac12\Big((\vec{\tilde x}_{12}\cdot\vec\partial_{x_{12}})\vec x_{12} -(d+1)\,\vec{\tilde x}_{12}\Big)\,\delta(\vec x_{12})\nn\\ 
&\quad=-\frac{d+1}{2}\,\vec{\tilde x}_{12}\,\delta(\vec x_{12})
\end{align}
In a Carrollian CFT in $d+1$ dimensions invariant under the Carrollian conformal algebra, the two point function between two scalar fields with the same scaling dimension $\Delta$ is
\begin{align}\label{2ptCCA}
\Delta_1=\Delta_2=\Delta:\quad    \boxed{G^{(2)}(\vec x_1,t_1;\vec x_2,t_2)=   C_1\, |\vec x_{12}|^{-2\Delta}+C_2\, |t_{12}|^{\frac{d-2\Delta}{z}}\delta(\vec x_{12})}\,.
\end{align}
which agrees for $z=1$ and $d=3$ with \cite{Chen:2021xkw}. We especially notice the presence of the ultra-local term which is not obtained in a naive way from the Carrollian contraction $\vec x\mapsto \vec x$ and $t\mapsto\epsilon \,t$ with $\epsilon\to0$ of the relativistic CFT two-point function when $\vec x_1\neq \vec x_2$
\begin{align}\label{relat2ptfct}
    \langle\phi(t_1,\vec x_1)\phi(t_2,\vec x_2)\rangle\sim\left(-|t_{12}|^2+|\vec x_{12}|^2\right)^{-\Delta}\,.
\end{align}
Of course, for $\vec x_1=\vec x_2$ this is more subtle and there may be a sense in which the ultra-local term can be obtained as a Carrollian limit of a suitable regularisation (in the sense of distributions) of the relativistic two-point function \eqref{relat2ptfct}. We do not investigate this question.

More generally, the two-point function for two scalar fields of distinct scaling dimension $\Delta_1\neq \Delta_2$ in Carrollian CFT takes the form
\begin{align}\label{Carrolian2points}
\Delta_1\neq\Delta_2:\quad	\boxed{G^{(2)}(\vec x_{12}, t_{12})=C_2\, |t_{12}|^{\frac{d-(\Delta_1+\Delta_2)}{z}}\delta(\vec x_{12})}\,.
\end{align} 
This is to be contrasted with the two-point functions of a relativistic CFT which must vanish for distinct scaling dimensions.

The special cases $z=0$  and $z=\infty$  highlight distinct symmetry constraints under spatial SCT $K_i$  on the two-point function: 
\begin{itemize}
    \item 
 For $z=0$, the spatial SCT generator is given by $K_i = 2x_i (x_j \partial_j + \Delta) - x^2 \partial_i$. Starting from equation \eqref{Carrolian2points'}, the symmetry constraint leads to $\Delta_1 = \Delta_2$, aligning the scaling dimensions of the two operators. However, there is no additional restriction on the functional form of $F(t_{12})$, which remains arbitrary. This reflects the residual freedom in the time dependence of the two-point function under Carrollian symmetry for $z=0$. 

\item For $z=\infty$, the spatial SCT generator is given by $K_i = 2x_i (t \partial_t+\Delta)$. As discussed by Duval et al \cite{Duval:2014lpa}, the finite spatial SCTs for $z=\infty$ should be thought as space-dependent time dilatation $t'= f(\vec x)\, t$ where $f$ is any nowhere-vanishing function on space. In equation \eqref{Carrolian2points'''} we have no restriction on $C_2$ for any value of $\Delta_1 + \Delta_2$. However, when $\Delta_1 + \Delta_2=0$ we have either $\Delta_1 = \Delta_2=0$ and $G(\vec x_{12})$ is not constrained, or we have $\Delta_1\neq 0$ or $\Delta_2\neq 0$ and $G(\vec x_{12})=0$.
\end{itemize}
These cases demonstrate how the specific forms of $K_i$ at $z=0$ and $z=\infty$ result in distinct symmetry constraints and freedoms for the two-point correlators under Carrollian scaling.

\subsubsection{Infinite-dimensional algebras: automatic enhancement}

One should finally note that the two-point functions \eqref{2ptCCA} and \eqref{Carrolian2points} happen to be invariant under the whole infinite-dimensional conformal Carroll algebra for any $z$. 
This can be easily checked by means of Carrollian conformal primary fields $\phi(t,\vec x)$ and the way they transform under conformal Carroll transformations. 

A conformal Carroll primary field of scaling dimension $\Delta$ is defined by the transformation law (see e.g. Section 2.2 in \cite{Bekaert:2022ipg})
\begin{align}\label{primarydef}
    \phi(\vec x, t)\to \phi'(\vec{x}\,',t')=\left|\frac{\partial \vec{x}\,'}{\partial \vec{x}}\right|^{-\Delta/d}\phi(\vec x, t)\,,
\end{align}
where 
\begin{equation}
    \left|\frac{\partial \vec{x}\,'}{\partial \vec{x}}\right|=\left|\det\left(\frac{\partial x'^{\,i}}{\partial x^j}\right)\right|
\end{equation}
denotes the absolute value of the Jacobian of the corresponding conformal transformation $x^i\to  x'^i(x^j)$ of the spatial coordinates.
The factor in the definition \eqref{primarydef} does not depend on the dynamical exponent $z$ because what dictates the scaling behaviour of a primary field is the conformal factor $\Omega(\vec x)=\left|\frac{\partial \vec{x}\,'}{\partial \vec{x}}\right|^{1/d}$ taken by the Carrollian metric, i.e. $d\vec x\,^{'2}= \Omega^2(\vec x)\,d\vec x^{2}$. Note that this conformal factor is independent of time. This makes clear why the magnetic part of the two-point function \eqref{2ptCCA} is actually invariant by itself.

A Carrollian conformal primary scalar fields of scaling dimension $\Delta$ picks a factor $\Omega(x)^{-\Delta}$ while the Carrollian time coordinate $t$ picks a factor $\Omega(x)^z$.
Therefore, to prove that the two-point function \eqref{Carrolian2points} is invariant and that the electric part of the two-point function \eqref{2ptCCA} is invariant by itself, one should simply realise that the Dirac delta function $\delta(\vec x)$ picks a factor $\Omega(x)^{-d}$. It is then clear that \eqref{Carrolian2points} is invariant. The latter transformation law of the Dirac delta function follows from the fact that, for any invertible coordinate transformation $x^i\to x'^{\,i}(x^j)$, we have the property
\begin{equation}\label{deltaJ}
    \delta\big(\vec x\,'(\vec x)-\vec y\,'(\vec y)\big)=\frac1{J}\,\delta(\vec x-\vec y)\,,\quad\text{with}\quad J= \left|\frac{\partial \vec{x}\,'}{\partial \vec{x}}\right|\,,
\end{equation}
where $\vec x\,'(\vec x)=\vec y\,'(\vec y)$ iff $\vec x=\vec y$ (for the considered invertible coordinate change). Note that $\vec x$ is taken as the variable in the formula above, hence $\vec y$ and $\vec y\,'$ are taken as constant vectors.
The identity \eqref{deltaJ} implies
\begin{equation}
    \delta(\vec x'_{12})=\delta\big(\vec x'_1(\vec x_1)-\vec x'_2(\vec x_2)\big)=\frac1{J}\,\delta(\vec x_1-\vec x_2)=\frac1{J}\,\delta(\vec x_{12})\,,\quad\text{with}\quad J=\left|\det\left(\frac{\partial x_1^{\prime\,i}}{\partial x_1^j}\right)\right|_{\vec x_1=\vec x_2}\,,
\end{equation}
where we used the fact that the coordinate transformation is bijective, in particular $\vec x'_1(\vec x_1)=\vec x'_2(\vec x_2)$ iff $\vec x_1=\vec x_2$. This ends the proof that the ultra-local terms are invariant under the infinite-dimensional groups of conformal Carroll transformations including supertranslations. The magnetic term in the two-point function is manifestly invariant also under these symmetries since supertranslations act trivially on this term and since it transforms properly under spatial conformal transformations.

In this way, we have shown that there is an enhancement of symmetries for all two-point functions which are invariant under a finite-dimensional conformal Carroll algebra of type D-K with generic dynamical exponent $z$ (cf. Section \ref{classification}) to the corresponding infinite-dimensional conformal Carroll algebra (cf. Section \ref{infdim}). In particular, an important corollary for $z=1$ is that any free QFT invariant under the Carrollian conformal algebra is also invariant under the full BMS algebra.

\subsection{Three-point functions}

If we consider the three-point function
\begin{align}
    G^{(3)}(\vec x_1, \vec x_2, \vec x_3; t_1, t_2, t_3):=\langle0|\phi_1(t_1,\vec x_1)\phi_2(t_2,\vec x_2)\phi_3(t_3,\vec x_3)|0\rangle
\end{align}
Applying the translation invariance in \eqref{Qinv} for $n=3$ we find that the three-point function must be a function of the differences $t_{ij}$ and $\vec x_{ij}$ ($i\neq j$). Thus we get $G^{(3)}=G(\vec x_{12},\vec x_{23}, \vec x_{31}; t_{12}, t_{23}, t_{31})$ where $t_{ij}=t_i-t_j$ and $\vec x_{ij}=\vec x_i-\vec x_j$. We notice that among the three differences only two of them are independent, since $t_{31}=-t_{12}-t_{23}$ and $\vec x_{31}=-\vec x_{12}-\vec x_{23}$. Nevertheless in order to look more symmetric, we are free to make the dependence of the 3-pt function on $t_{31}$ and $\vec x_{31}$ explicit as well. 
If we consider $Q$ in \eqref{Qinv} as the generator of the Carrollian boost $\vec B =\vec x \,\partial_t$ we have 
\begin{align}
 0&=  (\vec x_1\partial_{t_1}+\vec x_2\partial_{t_2}+\vec x_3\partial_{t_3})G^{(3)}(\vec x_{12}, \vec x_{23}, \vec x_{31}; t_{12}, t_{23}, t_{31})\nn\\
 &=(\vec x_{12}\partial_{t_{12}}+\vec x_{23}\partial_{t_{23}})G^{(3)}(\vec x_{12}, \vec x_{23}, \vec x_{31}; t_{12}, t_{23}, t_{31})\,,\label{constraintCarrollboost}
\end{align}
which is solved as, 
\begin{equation}\label{Carrolian3pointpre}
\boxed{~~
\begin{aligned}
&G^{(3)}(\vec x_{12}, \vec x_{23}, \vec x_{31}; t_{12}, t_{23}, t_{31})=G\big(|\vec x_{12}|,|\vec x_{23}|\big)+F(t_{12}, t_{23})\,\delta (\vec x_{12})\,\delta (\vec x_{23})\\[8pt]
 &\quad+F_1\big(|\vec x_{12}|;t_{23}\big)\,\delta (\vec x_{23})+F_2\big(|\vec x_{23}|;t_{31}\big)\,\delta (\vec x_{31})+F_3\big(|\vec x_{31}|;t_{12}\big)\,\delta (\vec x_{12})\\[8pt]
 &\quad+ \mathcal{F}_1\left(|\vec x_{12}|,|\vec x_{23}|\,;\, \frac{t_{12}}{|\vec x_{12}|}-\frac{t_{23}}{|\vec x_{23}|}\right)\,
 \delta\left(\frac{\vec x_{12}}{|\vec x_{12}|}-\frac{\vec x_{23}}{|\vec x_{23}|}\right)\\[3pt]
 &\quad+\mathcal{F}_2\left(|\vec x_{23}|,|\vec x_{31}|\,;\,\frac{t_{23}}{|\vec x_{23}|}-\frac{t_{31}}{|\vec x_{31}|}\right)\,\delta\left(\frac{\vec x_{23}}{|\vec x_{23}|}-\frac{\vec x_{31}}{|\vec x_{31}|}\right)\\[3pt]
 &\quad+ \mathcal{F}_3\left(|\vec x_{31}|,|\vec x_{12}|\,;\, \frac{t_{31}}{|\vec x_{31}|}-\frac{t_{12}}{|\vec x_{12}|}\right)\,\delta\left(\frac{\vec x_{31}}{|\vec x_{31}|}-\frac{\vec x_{12}}{|\vec x_{12}|}\right)\,,
\end{aligned}
~~}
\end{equation}
where we have written only the independent variables of the functions for compactness.
The first term on the right-hand side  of \eqref{Carrolian3pointpre} is of magnetic type, while the remaining terms are electric: the second term on the right-hand side and the three terms on the second line are ultra-local (like in the two-point function) while the last three terms on the right-hand side have a larger support. The Dirac delta functions in the second term on the right-hand side impose that the three points coincide: $\vec x_1=\vec x_2=\vec x_3$. In the three terms on the second line, they impose that two points coincide. And for the last three terms on the right-hand side, the Dirac delta functions basically enforce the difference vectors $\vec x_{12}$, $\vec x_{23}$, $\vec x_{31}$ to be all collinear (since $\vec x_{12}+\vec x_{23}+\vec x_{31}=0$), hence the three points $\vec x_1$, $\vec x_2$ and $\vec x_3$ must be aligned, as one can easily visualize:
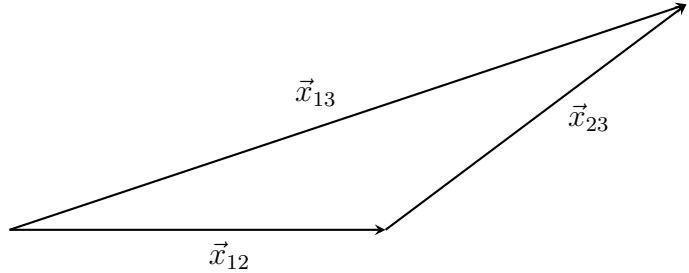
\begin{center}
\begin{tikzpicture}
\coordinate (O) at (0,0);
\coordinate (A) at (5,0);
\coordinate (B) at (9,3);

\draw[->, thick, >=stealth ] (O) -- (A) node[midway, below right] {$\vec{x}_{12}$};
\draw[->, thick, >=stealth ] (A) -- (B) node[midway, right] {$~~\vec{x}_{23}$};
\draw[-> , thick, >=stealth] (O) -- (B) node[midway, above left] {$\vec{x}_{13}$};
\end{tikzpicture}
\end{center}
There are 3 independent cases of such collinearity, depending on which point is in the middle, so there are 3 independent such terms as in \eqref{Carrolian3pointpre}. Note that for $d=1$, the collinearity is automatic and, thus, the these last three terms are only present for $d>1$. 

To prove that \eqref{Carrolian3pointpre} is the most general solution to \eqref{constraintCarrollboost}, we can rewrite the equation in three different ways to have manifest symmetry under circular permutations of the 3 points
\begin{subequations}\label{seperate3boost}
\begin{align}
    (\vec x_{12}\partial_{t_{12}}+\vec x_{23}\partial_{t_{23}})G^{(3)}&=0\,,\\
 (\vec x_{23}\partial_{t_{23}}+\vec x_{31}\partial_{t_{31}})G^{(3)}&=0\,,\\
    ( \vec x_{31}\partial_{t_{31}}+\vec x_{12}\partial_{t_{12}})G^{(3)}&=0  
\,.
\end{align} 
\end{subequations}
Let us make a change of variables as
\begin{align}
u_1=t_{23}-\frac{\vec x_{12}\cdot \vec x_{23}}{|\vec x_{12}|^2}\,t_{12}\,,\;\quad u_2=t_{31}-\frac{\vec x_{23}\cdot \vec x_{31}}{|\vec x_{23}|^2}\,t_{23}\,,\;\quad u_3=t_{12}-\frac{\vec x_{31}\cdot \vec x_{11}}{|\vec x_{31}|^2}\,t_{31}\,,
\end{align}
where only two out of those three variables are independent. After this change of variables, ${G}^{(3)}=\mathcal{G}(\vec x_{12}, \vec x_{23}, \vec x_{31}; u_1,u_2,u_3)$ and eqs. \eqref{seperate3boost} turn to
\begin{subequations}\label{seperate3boost12}
\begin{align}
    \vec x_{12}\neq 0:\quad\Big((\vec x_{12}\cdot\vec x_{23})\,\vec x_{12}-|\vec x_{12}|^2\,\vec x_{23}\Big)\partial_{u_1}\mathcal {G}&=0\,,\\
   \vec x_{23}\neq 0:\quad \Big((\vec x_{23}\cdot\vec x_{31})\,\vec x_{23}-|\vec x_{23}|^2\,\vec x_{31}\Big)\partial_{u_2}\mathcal {G}&=0\,,\\
  \vec x_{31}\neq 0:\quad  \Big((\vec x_{31}\cdot\vec x_{12})\,\vec x_{31}-|\vec x_{31}|^2\,\vec x_{12}\Big)\partial_{u_3}\mathcal {G}&=0\,.
\end{align}
\end{subequations}
It is obvious that in $d=1$ these equations are trivially satisfied. For $d>1$, the general solution to \eqref{seperate3boost12}  can be written as
\begin{align}\label{444}
\mathcal {G}={G}(\vec x_{12}, \vec x_{23}, \vec x_{31})&+ \mathcal{G}_1\left(\vec x_{12}, \vec x_{23}, \vec x_{31}\,;\, u_1\right)\,\delta\Big((\vec x_{12}\cdot\vec x_{23})\,\vec x_{12}-|\vec x_{12}|^2\,\vec x_{23}\Big)\nn\\
&+\mathcal{G}_2\left(\vec x_{12}, \vec x_{23}, \vec x_{31}\,;\, u_2\right)\,\delta\Big((\vec x_{23}\cdot\vec x_{31})\,\vec x_{23}-|\vec x_{23}|^2\,\vec x_{31}\Big)\nn\\ 
&+\mathcal{G}_3\left(\vec x_{12}, \vec x_{23}, \vec x_{31}\,;\, u_3\right)\,\delta\Big((\vec x_{31}\cdot\vec x_{12})\,\vec x_{31}-|\vec x_{31}|^2\,\vec x_{12}\Big)\,.
\end{align}
In each of the last three terms above, two out of the three vectors $x_{12}$, $x_{23}$ and $x_{31}$ have to be collinear. However, the fact that $\vec x_{12}+\vec x_{23}+\vec x_{31}=0$ implies that all three of them are actually collinear. Thence, asumming $\vec x_{ij}\neq 0$ for all $i\neq j$, there is a unit $\vec n$ vector such that 
\begin{equation}\label{unitn}
\vec x_{ij}\neq 0:\quad  \vec n=\epsilon_3\,\frac{\vec x_{12}}{|\vec x_{12}|}=\epsilon_1\,\frac{\vec x_{23}}{|\vec x_{23}|}=\epsilon_2\,\frac{\vec x_{31}}{|\vec x_{31}|}
\end{equation}
where $\epsilon_i =\pm 1$.
Consequently, for $\vec x_{ij}\neq\vec 0$ one can take into account all possible ways of being collinear and rearrange the last three terms on the right-hand side of \eqref{444} as the last three terms in \eqref{Carrolian3pointpre}. One gets the three terms in the third line of \eqref{Carrolian3pointpre} via the convenient rewriting
\begin{align}\label{445}
&&\vec x_{ij}\neq 0:\quad\mathcal {G}={G}(\vec x_{12}, \vec x_{23}, \vec x_{31})+ \mathcal{F}_1\left(\vec x_{12}, \vec x_{23}, \vec x_{31}\,;\, v_1\right)\,\delta\left(\frac{\vec x_{12}}{|\vec x_{12}|}-\frac{\vec x_{23}}{|\vec x_{23}|}\right)\nn\\
&&\qquad+\mathcal{F}_2\left(\vec x_{12}, \vec x_{23}, \vec x_{31}\,;\, v_2\right)\,\delta\left(\frac{\vec x_{23}}{|\vec x_{23}|}-\frac{\vec x_{31}}{|\vec x_{31}|}\right)+\mathcal{F}_3\left(\vec x_{12}, \vec x_{23}, \vec x_{31}\,;\, v_3\right)\,\delta\left(\frac{\vec x_{23}}{|\vec x_{23}|}-\frac{\vec x_{31}}{|\vec x_{31}|}\right)
\end{align}
and change of variables
\begin{equation}\label{Carrolian3pointpre11pr}
    v_1=\frac{t_{12}}{|\vec x_{12}|}-\frac{t_{23}}{|\vec x_{23}|}\,,\quad v_2=\frac{t_{23}}{|\vec x_{23}|}-\frac{t_{31}}{|\vec x_{31}|}\,,\quad v_3=\frac{t_{31}}{|\vec x_{31}|}-\frac{t_{12}}{|\vec x_{12}|}\,.
\end{equation}
Note that there is a complicated relation between the functions $\mathcal{G}_i$ in \eqref{444} and the functions $\mathcal{F}_i$ in \eqref{Carrolian3pointpre11pr} but there is no loss of generality in this convenient rewriting. To see this, consider for instance the case $\epsilon_1=1=\epsilon_3$ in \eqref{unitn}. Then the condition $\vec x_{12}+\vec x_{23}+\vec x_{31}=0$ imposes that $\epsilon_2=-1$. One can then check that each Dirac delta function in \eqref{444} becomes proportional to $\delta\big(\frac{\vec x_{12}}{|\vec x_{12}|}-\frac{\vec x_{23}}{|\vec x_{23}|}\big)$, e.g. one has
$\delta\big((\vec x_{12}\cdot\vec x_{23})\,\vec x_{12}-|\vec x_{12}|^2\,\vec x_{23}\big)=
\frac1{|\vec x_{23}|^d}\delta\big(\frac{\vec x_{12}}{|\vec x_{12}|}-\frac{\vec x_{23}}{|\vec x_{23}|}\big)$ and $t_{23}-\frac{\vec x_{12}\cdot \vec x_{23}}{|\vec x_{12}|^2}\,t_{12}=|\vec x_{23}|\big(\frac{t_{23}}{|\vec x_{23}|}-\frac{t_{12}}{|\vec x_{12}|}\big)$. 
Finally, one should consider the remaining cases when one or more $\vec x_{ij}=\vec 0$. 
If either $\vec x_{12}=\vec 0$ or $\vec x_{23}=\vec 0$ or $\vec x_{31}=\vec 0$, then one gets the ultra-local terms in the first and second line of \eqref{Carrolian3pointpre} depending on how many points coincide.
This ends the proof of the general expression \eqref{Carrolian3pointpre}
for the three-point function in a Carrollian QFT.

Note that the magnetic term in \eqref{Carrolian3pointpre}, as well as the ultra-local terms, admit an infinite-dimensional enhancement of symmetries since they are invariant under any supertranslation $t\to t+f(\vec x)$, like for the two-point function. However, for three-point functions the electric collinear terms are not invariant under supertranslations, except if they are actually magnetic (i.e. if the functions $\mathcal{F}_i$ do not depend on $t_{ij}$) and can therefore be reabsorbed inside $G\big(|\vec x_{12}|,|\vec x_{23}|,|\vec x_{31}|\big)$. In fact, genuine electric collinear terms appear to be somewhat exotic: as will be seen below, the invariance of the collinear terms under temporal special conformal transformations is enough to remove this possibility.

\subsubsection{Dilatation invariance}

We can now consider instead the invariance under infinitesimal dilatations
\begin{align}\label{diff3pt}
 0&=\le(z\, t_1\partial_{t_1}+\vec x_1\cdot\vec\partial_{x_1}+z\, t_2\partial_{t_2}+\vec x_2\cdot\vec\partial_{x_2}+z\, t_3\partial_{t_3}+\vec x_3\cdot\vec\partial_{x_3}+\Delta_1+\Delta_2+\Delta_3\ri)G^{(3)}\nn\\[5pt]
 &=\le(z\, t_{12}\partial_{t_{12}}+\vec x_{12}\cdot\vec\partial_{x_{12}}+z\, t_{32}\partial_{t_{32}}+\vec x_{32}\cdot\vec\partial_{x_{32}}+\Delta_1+\Delta_2+\Delta_3\ri)G^{(3)}\,.
 \end{align}
It turns out to be simpler to consider instead the covariance of the three-point function \eqref{Carrolian3pointpre} under finite dilatations. Note that the various types of terms will not mix under such transformations. In other word, each term in \eqref{Carrolian3pointpre} will have to transform as the left-hand side under dilatations. The conformal dimension $\Delta$ for fields transforming under scaling $t\to\lambda^zt$ and ${\vec x}\to\lambda \vec x$ is defined according to
\begin{align}
	\phi(t,{\vec x})\to
 \lambda^{\Delta}\phi(\lambda^zt,\lambda {\vec x})\,.
\end{align}
This implies that the magnetic piece $G\big(|\vec x_{12}|,|\vec x_{23}|,|\vec x_{31}|\big)$ is a homogeneous function of two independent variables, say $|\vec x_{12}|$ and $|\vec x_{23}|$, with degree $-(\Delta_1+\Delta_2+\Delta_3)$. Therefore, the homogeneous equation is solved as (cf. \cite[Subsection 4.3.1]{DiFrancesco:1997nk})
\begin{align}\label{3ptnodelta}
\boxed{G\big(|\vec x_{12}|,|\vec x_{23}|,|\vec x_{31}|\big)=\sum_{a+b+c=\Delta_1+\Delta_2+\Delta_3}\frac{C_{abc}}{|\vec x_{12}|^a|\vec x_{23}|^b|\vec x_{31}|^c}}
\end{align}
where $C_{abc}$ are arbitrary constants.

Similarly, the coefficient of the electric term supported when all three points coincide  must obey the homogeneity condition 
\begin{equation}\label{FDKabc}
    F(\lambda^z t_{12},\lambda^z t_{23},\lambda^z t_{31})=\lambda^{2d-(\Delta_1+\Delta_2+\Delta_3)}F(t_{12}, t_{23}, t_{31})\,,\qquad\forall \lambda>0\,,
\end{equation}
which can be solved as
\begin{align}\label{3ptnodeltas}
z\neq 0:\quad   \boxed{F(t_{12},t_{23},t_{31})=\sum_{a+b+c\,=\,\frac{\Delta_1+\Delta_2+\Delta_3-2d}{z}}\,\frac{B_{abc}}{|t_{12}|^a|t_{23}|^b|t_{31}|^c}}\,,
\end{align}
where $B_{abc}$ are arbitrary constants.

We can proceed along similar lines for the remaining ultra-local terms and get for instance 
\begin{align}\label{F_1dilatinv}
z\neq 0:\quad  \boxed{F_1\big(|\vec x_{12}|,t_{23}\big)=\sum_{a+b=\Delta_1+\Delta_2+\Delta_3-d}\frac{K_{ab}}{|\vec x_{12}|^{a}|t_{23}|^{\frac{b}{z}}}}\,,
\end{align}
where $K_{ab}$ are arbitrary constants, and similarly for $F_2$ and $F_3$.
Finally, for the collinear terms the argument of the Dirac delta functions are scale invariant, therefore one gets for instance
\begin{equation}\label{mathcalF1}
z\neq 1:\quad \boxed{\mathcal{F}_1\big(|\vec x_{12}|,|\vec x_{23}|\,;\, v_1\big)=\sum_{a+b+c=\Delta_1+\Delta_2+\Delta_3}\frac{\mathcal{K}_{abc}}{|\vec x_{12}|^a|\vec x_{23}|^b|v_1|^{\frac{c}{z-1}}}}\,,
\end{equation}
where $\mathcal{K}_{abc}$ are arbitrary constants, and similarly for $\mathcal{F}_2$ and $\mathcal{F}_3$ (via circular permutations of the indices). For $z=1$, the variable $v_1$ in \eqref{Carrolian3pointpre11pr} is dilatation invariant, therefore the only condition that one gets instead is
\begin{equation}
z=1:\quad \mathcal{F}_i\big(\lambda|\vec x_{12}|,\lambda|\vec x_{23}|,\lambda|\vec x_{31}|\,;\, v_i\big)=\lambda^{-(\Delta_1+\Delta_2+\Delta_3)} \mathcal{F}_i\big(|\vec x_{12}|,|\vec x_{23}|,|\vec x_{31}|\,;\, v_i\big)\,.
\end{equation}
Therefore,
\begin{equation}
z=1:\quad   \boxed{\mathcal{F}_i\big(|\vec x_{12}|,|\vec x_{23}|,|\vec x_{31}|\,;\, v_i\big)=\sum_{a+b+c=\Delta_1+\Delta_2+\Delta_3}\frac{K^{(i)}_{abc}(v_i)}{|\vec x_{12}|^a|\vec x_{23}|^b|\vec x_{31}|^c}}
\end{equation}
where $K^{(i)}_{abc}(v_i)$ are arbitrary functions of the single variable $v_i$.

\subsubsection{K-invariance}\label{typeDK3pt}

After considering the dilatation generator $D$, let us have a look at the invariance of the 3-point function under  the temporal special conformal transformation generated by $K$;
\begin{align}
    0&=\le(\vec x_1\cdot\vec x_1\,\partial_{t_1}+\vec x_2\cdot \vec x_2\,\partial_{t_2}+\vec x_3\cdot \vec x_3\,\partial_{t_3}\ri)G^{(3)}(\vec x_{12}, \vec x_{23}, \vec x_{31}; t_{12}, t_{23}, t_{31})\nn\\[5pt]
    &=\big[(\vec x_1\cdot\vec x_1-\vec x_2\cdot \vec x_2)\partial_{t_{12}}+(\vec x_3\cdot \vec x_3-\vec x_2\cdot \vec x_2)\partial_{t_{32}}\big]G^{(3)}(\vec x_{12}, \vec x_{23}, \vec x_{31}; t_{12}, t_{23}, t_{31})
\end{align}
which is trivially satisfied by the magnetic piece, but also  automatically satisfied by the electric ultra-local pieces. This is clear since, as mentioned above, these terms are invariant under all supertranslations (remember that the generator $K$ is a particular example of supertranslation generator).

Now, let us examine the invariance of the collinear terms in the 3-point function under the temporal special conformal transformation. The invariance under the generator $K$ gives, for each collinear term, a condition similar to this one
\begin{equation}
    \left(\frac{\vec x_{12}\cdot (\vec x_1+\vec x_2)}{|\vec x_{12}|}-\frac{\vec x_{23}\cdot (\vec x_2+\vec x_3)}{|\vec x_{23}|}\right)\partial_{v_1}\mathcal F_1\,\delta\left(\frac{\vec x_{12}}{|\vec x_{12}|}-\frac{\vec x_{23}}{|\vec x_{23}|}\right)=0
\end{equation}
\begin{equation}
\implies  \frac{\vec x_{12}\cdot \vec x_{13}}{|\vec x_{12}|}\,\partial_{v_1}\mathcal F_1\,\delta\left(\frac{\vec x_{12}}{|\vec x_{12}|}-\frac{\vec x_{23}}{|\vec x_{23}|}\right)=0
\end{equation}   
Therefore, either $\partial_{v_1}\mathcal F_1=0$ or $\vec x_{12}\cdot \vec x_{13}=0$. The latter is not possible since $\vec x_{13}=\vec x_{12}+\vec x_{23}$ and $\vec x_{12}=\frac{|\vec x_{12}|}{|\vec x_{23}|}\,\vec x_{23}$.
Therefore, $\partial_{v_i}\mathcal F_i=0$ ($i=1,2,3$) for the collinear terms. In other words, the collinear terms can be reabsorbed in the magnetic term. In summary, the $K$ invariance removes all collinear terms as independent solutions.
To conclude, the general form of the 3-point function covariant under the Carroll transformations as well as the temporal conformal transformations reads
\begin{equation}\label{Carrolian3pointpreK}
\boxed{~~
\begin{aligned}
&G^{(3)}(\vec x_{12}, \vec x_{23}, \vec x_{31}; t_{12}, t_{23}, t_{31})=G\big(|\vec x_{12}|,|\vec x_{23}|\big)+F(t_{12}, t_{23})\,\delta (\vec x_{12})\,\delta (\vec x_{23})\\[8pt]
 &\quad+F_1\big(|\vec x_{12}|;t_{23}\big)\,\delta (\vec x_{23})+F_2\big(|\vec x_{23}|;t_{31}\big)\,\delta (\vec x_{31})+F_3\big(|\vec x_{31}|;t_{12}\big)\,\delta (\vec x_{12})
\end{aligned}
~~}
\end{equation}

To conclude, the general form of the 3-point function covariant under the conformal Carroll algebra of type D-K is \eqref{Carrolian3pointpreK} where the coefficients are given by \eqref{3ptnodelta}, \eqref{3ptnodeltas}, \eqref{mathcalF1}, etc.

\subsubsection{K\texorpdfstring{$_i\,$}{}-invariance}

If we also impose the invariance under the spatial SCT's generated by $K_i$, we have
\begin{align}\label{SCTinv3pt}
&\Big(~\,\vec x_1\,(z\,t_1\partial_{t_1}+\vec x_1\cdot\vec\partial_{x_1})-\frac12(\vec x_1\cdot \vec x_1)\,\vec\partial_{x_1}+\Delta_1\,\vec x_1 \nn\\[5pt]&+	\vec x_2\,(z\,t_2\partial_{t_2}+\vec x_2\cdot\vec\partial_{x_2})-\frac12(\vec x_2\cdot \vec x_2)\,\vec\partial_{x_2}+\Delta_2\,\vec x_2\nn\\[5pt]&+	\vec x_3\,(z\,t_3\partial_{t_3}+\vec x_3\cdot\vec\partial_{x_3})-\frac12(\vec x_3\cdot \vec x_3)\,\vec\partial_{x_3}+\Delta_3\,\vec x_3\Big)\,G^{(3)}=0\,.    
\end{align}
where $G^{(3)}$ is given in \eqref{Carrolian3pointpreK}. Again let us consider the case where $x_{ij}\neq0$ and thus all ultra-local terms are absent, then the action of the time derivative in the generator ${ K_i}$ is trivial and following the standard arguments (see e.g. \cite[Subsection 4.3.1]{DiFrancesco:1997nk}), we should have the following constraints on the magnetic part of the three-point function 
\begin{align}\label{abcDelta}
    a+c=2\Delta_1\,,\qquad a+b=2\Delta_2\,,\qquad b+c=2\Delta_3\,.
\end{align}
The form of the spatial (magnetic) part  of the 3-pt $G^{(3)}$ becomes
\begin{equation}\label{G3ptinvmag}
\boxed{~~
\begin{aligned}
G(\vec x_{12}, \vec x_{23}, \vec x_{31})=\frac{C_1}{|\vec x_{12}|^{\Delta_1+\Delta_2-\Delta_3}|\vec x_{23}|^{\Delta_3+\Delta_2-\Delta_1}|\vec x_{31}|^{\Delta_3+\Delta_1-\Delta_2}}
\end{aligned}
~~}
\end{equation}
Now we intend to impose the spatial SCT on the electric part. For the first ultra-local term in \eqref{Carrolian3pointpreK} we have
\begin{align}\label{SCTinv3ptF}
&\Big[\vec x_1 (\vec x_1\cdot\vec\partial_{x_{12}}) -\vec x_2(\vec x_2\cdot\vec\partial_{x_{12}})+\vec x_2(\vec x_2\cdot \vec\partial_{x_{23}}) -\vec x_3(\vec x_3\cdot\vec\partial_{x_{23}})+z(\vec x_1t_1-\vec x_2t_2)\partial_{t_{12}}
+z(\vec x_2t_2-\vec x_3t_3)\partial_{t_{23}}\nn\\[6pt]
&\!\!\!-\frac12(\vec x_1\cdot\vec x_1-\vec x_2\cdot\vec x_2)\vec\partial_{x_{12}}-\frac12(\vec x_2\cdot\vec x_2-\vec x_3\cdot\vec x_3)\vec\partial_{x_{23}}+\vec x_1\Delta_1+\vec x_2\Delta_2+\vec x_3\Delta_3\Big]\,F(t_{12}, t_{23})\,\delta (\vec x_{12})\,\delta (\vec x_{23})\nn\\[6pt]
&\quad=0\,, 
\end{align}
where we ignored the $\vec x_{13}$ and $t_{13}$ dependence. If we use \eqref{3ptnodeltas} and  identities \eqref{Kiidentity1} and \eqref{Kiidentity2}  we have the condition
\begin{align}\label{SCTinv3ptF2}
F(t_{12}, t_{23})\Big[&-\frac{d+1}{2}\,\vec{\tilde x}_{12}-\frac{d+1}{2}\,\vec{\tilde x}_{32}-az(\vec x_1t_1-\vec x_2t_2)t_{12}^{-1}
-bz(\vec x_2t_2-\vec x_3t_3)t_{23}^{-1}-cz(\vec x_3t_3-\vec x_1t_1)t_{31}^{-1}
\nn\\[6pt]
&+\frac12\vec {\tilde x}_{12}+\frac12\vec {\tilde x}_{32}+\vec x_1\Delta_1+\vec x_2\Delta_2+\vec x_3\Delta_3\Big]\,\,\delta (\vec x_{12})\,\delta (\vec x_{23})=0\,.
\end{align}
Once we use the fact that in this term $\vec x_1=\vec x_2=\vec x_3$, we see that this relation is identically zero. If we impose the spatial SCT on ultra-local terms in the 2nd line of \eqref{Carrolian3pointpreK} we have the condition 
\begin{align}\label{SCTinv3ptF3}
&0=\Big[(\vec x_1\vec x_1-\vec x_2\vec x_2)\cdot\vec\partial_{x_{12}}+(\vec x_2\vec x_2-\vec x_3\vec x_3)\cdot\vec\partial_{x_{23}}
+z(\vec x_2t_2-\vec x_3t_3)\partial_{t_{23}}\\[6pt]
&\!\!-\frac12(\vec x_1\cdot\vec x_1-\vec x_2\cdot\vec x_2)\vec\partial_{x_{12}}-\frac12(\vec x_2\cdot\vec x_2-\vec x_3\cdot\vec x_3)\vec\partial_{x_{23}}+\Delta_1\,\vec x_1+\Delta_2\,\vec x_2+\Delta_3\,\vec x_3\Big]\,F_1(\vec x_{12},t_{23})\,\delta (\vec x_{23})\,. \nn
\end{align}
Applying the derivatives to \eqref{F_1dilatinv} gives
\begin{align}\label{SCTinv3ptF4}
&\Big[-a(\vec x_1\vec x_1-\vec x_2\vec x_2)\cdot{\vec x}_{12}\,|\vec x_{12}|^{-2}
-\frac{d+1}{2}\,\vec{\tilde x}_{32}
-b(\vec x_3t_3-\vec x_2t_2){t_{32}}^{-1}+\frac{a}{2}(\vec x_1\cdot\vec x_1-\vec x_2\cdot\vec x_2){\vec x}_{12}\,|\vec x_{12}|^{-2}
\nn\\[5pt]
&\quad+\frac12{\vec {\tilde x}_{32}}+\Delta_1\,\vec x_1+\Delta_2\,\vec x_2+\Delta_3\,\vec x_3\Big]\,F_1(\vec x_{12}, t_{23})\,\delta (\vec x_{23})\nn\\[5pt]
&=\Big[-a(\vec x_1\vec x_1-\vec x_2\vec x_2)\cdot{\vec x}_{12}\,|\vec x_{12}|^{-2}+\frac{a}{2}(\vec x_1\cdot\vec x_1-\vec x_2\cdot\vec x_2)\,{\vec x}_{12}\,|\vec x_{12}|^{-2}\nn\\[5pt]&+\Delta_1\,\vec x_1+(\Delta_2+\Delta_3-d-b)\,\vec x_2\Big]\,F_1(\vec x_{12},t_{23})\,\delta (\vec x_{23})\nn\\[5pt]
&=\left[-\frac{a}{2}\left(|\vec x_1|^2+|\vec x_2|^2-2\vec x_1\cdot\vec x_2\right)\vec{\tilde x}_{12}\,|\vec x_{12}|^{-2}+\Delta_1\,\vec x_1+(a-\Delta_1)\vec x_2\right]\,F_1(\vec x_{12}, t_{23})\,\delta (\vec x_{23})\nn\\[5pt]
&=\left[-\frac{a}{2}\,\vec{\tilde x}_{12}+\Delta_1\,\vec x_{12}+a\,\vec x_2\right]\,F_1(\vec x_{12}, t_{23})\,\delta (\vec x_{23})\nn\\[5pt]
&=\left(\Delta_1-\frac{a}{2}\right)\vec{x}_{12}\,F_1(\vec x_{12}, t_{23})\,\delta (\vec x_{23})\,.
\end{align}
where we  used  $\Delta_2+\Delta_3-d-b=a-\Delta_1$ in the second equality. The invariance of this term imposes $a=2\Delta_1$ we thus have,
\begin{equation}\label{coefsF1}
\boxed{~~
\begin{aligned}
    F_1\big(|\vec x_{12}|,t_{23}\big)&=\frac{K_1}{|\vec x_{12}|^{2\Delta_1}|t_{23}|^{\frac{\Delta_2+\Delta_3-\Delta_1-d}{z}}}\\[6pt]
    F_2\big(|\vec x_{23}|,t_{31}\big)&=\frac{K_2}{|\vec x_{23}|^{2\Delta_2}|t_{31}|^{\frac{\Delta_1+\Delta_3-\Delta_2-d}{z}}}\\[6pt]
    F_3\big(|\vec x_{31}|,t_{12}\big)&=\frac{K_3}{|\vec x_{31}|^{2\Delta_3}|t_{12}|^{\frac{\Delta_1+\Delta_2-\Delta_3-d}{z}}}
\end{aligned}
~~}
\end{equation}
where we applied the delta function as well e.g. $x_{12}=x_{13}$ in $F_1$.
We applied $K_i$-invariance on  the $F_3$ term as well.

We thus see that imposing the spatial SCT does  constrain the electric ultra-local terms as a result of $\vec x$-dependency of these terms. This is unlike the case of the two-point function. This feature is also in contrast with the fact that the temporal SCT did not constrain the magnetic part of the correlation function.

\section{Finite transformations and Carrollian inversions} \label{Carrollianinversion}

The invariance of correlators under the finite Carroll conformal transformations generated by \eqref{CCG} (with suitable coefficients) is somewhat easy to check for all transformations, except for the spatial SCT's. On the one hand, we observe that, for any $z$, the finite form of the transformations generated by $K_i=2\,x_i D - \,x^2\p_i$ with $D=z\,t\,\p_t+x^i\p_i$ are 
\begin{equation}\label{SCT}
    \vec x^{\,\prime}=\frac{\vec x-|\vec x|^2\,\vec b}{1-2\,\vec b\cdot\vec x+|\vec b|^2|\vec x|^2}\,,\qquad t^\prime=\frac{t}{\big(1-2\,\vec b\cdot\vec x+|\vec b|^2|\vec x|^2\big)^z}\,.
\end{equation}
These spatial SCT's form an additive abelian subgroup $\mathbb{R}^d$.
On the other hand, the transformation generated by $K= x^2\partial_t$ are\footnote{These transformation can be obtained via contraction $c\to0$ from the relativistic case when $z=1$;
\begin{align*}
    {x'}^\mu=\frac{x^{\mu}-b^\mu x^\nu x_{\nu}}{1-2b^\nu x_{\nu}+b^\nu b_{\nu}x^{\rho} x_{\rho}}\,.
\end{align*}
}
\begin{align}\label{tempSCT}
        \vec x^{\,\prime}=\vec x\,,\qquad t^\prime=t+\alpha \,|\vec x|^2 \,.
\end{align}
The special conformal transformations \eqref{SCT}-\eqref{tempSCT} are examples of finite conformal Carroll transformation in the geometric sense of \cite{Duval:2014lpa}.
A \textit{Carrollian inversion} will be defined as the involutive transformation
\begin{equation}\label{inversion}
    \vec x^{\,\prime}=\frac{\vec x}{\,|\vec x|^2}\,,\qquad t^\prime=\frac{t}{\,|\vec x|^{2z}}\,.
\end{equation}
It is also an example of finite conformal Carroll transformation in the sense of \cite{Duval:2014lpa}.
By rewriting the spatial SCT \eqref{SCT} equivalently as follows,
\begin{equation}\label{SCT'}
    \frac{\vec x^{\,\prime}}{\,|\vec x^{\,\prime}|^2}=\frac{\vec x}{\,|\vec x|^2}-\vec b\,,\qquad \frac{t^\prime}{\,|\vec x^{\,\prime}|^{2z}}=\frac{t}{\,|\vec x|^{2z}}\,,
\end{equation}
it becomes manifest that spatial SCT's are conjugated to spatial translations via a Carrollian inversion. Indeed, let us first
consider a spatial SCT, say the spacetime transformation $\exp(\vec b\cdot \vec K)$ acting as \eqref{SCT}. Second, let us consider a spatial translation, say $\exp(-\vec b\cdot \vec P)$ acting as $\vec x\mapsto \vec x-\vec b$ and $t\mapsto t$. Third, let us denote the Carrollian inversion as the spacetime transformation $\mathbb{I}$ defined by \eqref{inversion}. The spatial SCT is conjugated to the spatial translation via the Carrollian inversion in the sense that $\exp(\vec b\cdot \vec K)=\mathbb{I}\circ \exp(-\vec b\cdot \vec P)\circ\mathbb{I}$. This relation can be checked by direct computations but is essentially made manifest by the rewriting of the spatial SCT \eqref{SCT} as \eqref{SCT'}. 

The covariance of $n$-point functions under Carroll inversions using \eqref{primarydef} reads
\begin{equation}
    G^{(n)}\left(\frac{\vec x_1}{\,|\vec x_1|^2}, \cdots, \frac{\vec x_n}{\,|\vec x_n|^2};\frac{t_1}{\,|\vec x_1|^{2z}}, \cdots, \frac{t_n}{\,|\vec x_n|^{2z}}\right)=|\vec x_1|^{2\Delta_1}\cdots|\vec x_n|^{2\Delta_n}G^{(n)}(\vec x_1,\cdots, \vec x_n; t_1,\cdots, t_n)\,.
\end{equation}
An important consequence of the property that spatial SCT's are conjugated to spatial translations via a Carrollian inversion is that translation and inversion invariances provide a sufficient condition for the invariance under spatial SCT's. For instance, for any dynamical exponent $z\neq 0$ the two-point functions \eqref{2ptCCA} and \eqref{Carrolian2points} can be checked to be invariant under a Carrollian inversion. This follows from the following properties for the inversion \eqref{inversion}: $|\vec x_{ij}^{\,\prime}|=\frac{|\vec x_{ij}|}{|\vec x_i|\,|\vec x_j|}$ and $F(t^\prime_{ij})\delta(\vec x_{ij}^{\,\prime})=F\big(\frac{t_{ij}}{|\vec x_i|^{2z}}\big)|\vec x_i|^{2d}\,\delta(\vec x_{ij})$.
One can check similarly the invariance under inversions of three-point functions which are of the form   
\eqref{Carrolian3pointpreK} where the coefficients are respectively given by \eqref{G3ptinvmag} and \eqref{coefsF1}.

Another corollary is that, in any Carrollian QFT which is also invariant under dilatations with integer dynamical exponent $z=N\in\mathbb N$, the condition of supertranslation and inversion invariance is sufficient to guarantee that this Carrollian CFT is invariant under the corresponding algebra $\mathfrak{cca}_{N}(d+1)$.
Note that the conjugate of a temporal translation by an inversion defines a transformation such that
\begin{equation}\label{SCT''}
    \frac{\vec x^{\,\prime}}{\,|\vec x^{\,\prime}|^2}=\frac{\vec x}{\,|\vec x|^2}\,,\qquad \frac{t^\prime}{\,|\vec x^{\,\prime}|^{2z}}=\frac{t}{\,|\vec x|^{2z}}-a\,,
\end{equation}
which is the following supertranslation
\begin{equation}\label{SCT'''}
    \vec x^{\,\prime}=\vec x\,,\qquad t^\prime=t-a\,|\vec x|^{2z}\,.
\end{equation}
Such a supertranslation is a temporal translation iff $z=0$ and it is a temporal special conformal transformation \eqref{tempSCT} iff $z=1$. In this way, one can see that a Carrollian QFT invariant under inversions with $z=N>1$ must necessarily be invariant under some supertranslations beyond temporal special conformal transformations. This is similar with the observations about the $K_i$ invariance in Section \ref{infdim}. 

\section{Conclusions and outlook}\label{conclu}

In this work we have classified various conformal extensions of the Carroll algebra 
in terms of their dynamical exponent $z\in\mathbb{R}$ as the controller of the anisotropy scaling between space and time. We may summarize these extensions in  table \ref{Table2}.  We also determined the most general form of 2-point and 3-point correlators invariant under these conformal symmetries.
\begin{table}[h]
\begin{center}
	\begin{tabularx}{\textwidth}{|X|X|}
		\hline
		Carroll algebra $\oplus$ $\cdots$ & $\mathfrak{carr}(d+1)$ \\ \hline
	\end{tabularx}	

	\begin{tabularx}{\textwidth}{|X|}
 		\hline
  \vspace{-.6cm}
  \begin{center}
      finite-dimensional extensions: \vspace{-.3cm}
  \end{center}
		 \\ \hline
	\end{tabularx}
	\begin{tabularx}{\textwidth}{|X|X|}
		\hline 
		K &   $\mathfrak{Kcarr}(d+1)$  \\
		D$_z$ & $\mathfrak{scalcarr}_z(d+1)$  \\
		D$_z$~+~K  & $\mathfrak{confcarr}_z(d+1)$  \\
		D$_1$~+~K~+~K$_i$   & $\mathfrak{cca}_1(d+1)\cong\mathfrak{iso}(d+1,1)$  \\
		D$_{1/2}$~+~$\mathbb{K}$  & $\mathfrak{carrsch}(1+1)$  \\
		D$_{N/2}$~+~K~+~$\mathbb{K}$~+~some supertrans,~~ $N\geqslant 3$  & $\mathfrak{cca}_{N/2}(1+1)$   \\
		D$_{N}$~+~K~+~K$_i$~+~some supertrans,~~\,\,$N\geqslant 2$  & $\mathfrak{cca}_{N}(d+1)$ \qquad($d>1$)  \\
		\hline
	\end{tabularx}
	\begin{tabularx}{\textwidth}{|X|}
		\hline \vspace{-.6cm}\begin{center}
		    infinite-dimensional extensions: \vspace{-.3cm}
		\end{center}
  
		\\ \hline
	\end{tabularx}
	\begin{tabularx}{\textwidth}{|X|X|}
		\hline
		 supertranslations & $\mathfrak{isocarr}(d+1)$  \\
		 D$_1$~+~K~+~K$_i$~+~all supertranslations  & $\infcar_1(d+1)\cong\mathfrak{bms}_{d+2}$,~~\,\quad$d>2$  \\
		D$_1$~+~K~+~K$_i$~+~all super(trans/rot)  & $\Widetilde{\mathfrak{confcarr}}_1(d+1)\cong\mathfrak{ebms}_{d+2}$, \quad$d=1,2$ \\
		D$_z$~+~K~+~K$_i$~+~all super(trans/rot)  & $\Widetilde{\mathfrak{confcarr}}_z(d+1)$  \\
		\hline	
	\end{tabularx}
\end{center}
\caption{Summarize of all possible conformal extensions of the Carroll algebra and their acronyms.}\label{Table2}
\end{table}

Based on the findings presented in this study, several promising avenues for future research have emerged. These directions have the potential to expand the applications of the present work. The key areas for further investigation are as follows:

\paragraph{Flat-space holography.}
What are the consistent set of asymptotically locally flat boundary conditions? Although there exist some generalization of locally flat boundary conditions \cite{Campiglia_2014,Afshar:2015wjm,Donnay:2015abr,Afshar:2016wfy,Donnay:2016ejv,Grumiller:2019ygj,Grumiller:2017sjh,Afshar:2019axx,Grumiller:2019fmp}, a complete classification on the gravity side is missing. 

In the present article we tried to address this question on the other side of the duality by classifying all conformal extensions of the Carroll algebra in arbitrary spacetime dimensions, $d+1$, in terms of the anisotropic scaling exponent $z$ denoted as $\infcar_z(d+1)$ in \eqref{confcarrz}, \eqref{3dcca} and \eqref{i>4td}. The famous isomorphism \eqref{bmscca1} in our notation 
\begin{align}\label{bmscca}
\infcar_1(d+1) \cong \mathfrak{bms}_{d+2}\,.
\end{align}
corresponds to $z=1$ and we have a similar enhancement of symmetries associated to supertranslation (and, possibly, superrotation) for arbitrary $z$. For example, the $z=0$ conformal extension of the Carroll algebra which is called here ``spatial conformal Carroll algebra'', has a  holographic realization corresponding to the flat Rindler (non-extremal near-horizon) asymptotic symmetries in three, four and higher bulk dimensions which is a warped Virasoro algebra with vanishing U(1) level \eqref{2ditk} at the boundary \cite{Afshar:2015wjm,Donnay:2015abr,Grumiller:2019fmp};
\begin{align}\label{schcarr}
\infcar_0(d+1) \cong \mathfrak{Rindler}_{d+2}
\end{align}
Although there already exists holographic realization of these symmetries in asymptotically locally flat spacetimes in three and four dimensions at $z=1$ \cite{Bondi:1962px,Sachs:1962zza,Barnich:2006av,Barnich:2010eb,Troessaert:2017jcm,Henneaux:2018cst} and at $z=0$ \cite{Afshar:2015wjm,Donnay:2015abr,Afshar:2016wfy,Donnay:2016ejv,Grumiller:2019fmp}, it is very suggestive to setup consistent asymptotically locally flat boundary conditions for general $z$.

\paragraph{Infinite-dimensional conformal algebras.}
The conformal Carroll symmetries have an infinite-dimensional enhancement which is a particularly important feature, relevant to flat space holography discussed above. This enhancement basically occurs in the time shift (``supertranslation'') symmetry in any dimension. But, in lower (two and three) spacetime dimensions one also has enhancement in the spatial (``super-rotation'') sector, beyond spatial translations, dilatations and spatial SCT's. We discussed these cases thoroughly in section \ref{infdim}. They all have a semi-direct product structure of conformal symmetries of the $d$-sphere under which the supertranslations transform as primary fields with the dynamical exponent $z$ playing the role of their conformal weight.  Especially when $z$ is integer or half-integer bigger than one, there is a finite-dimensional truncation of these algebras, which we called ``extended Carrollian conformal algebra'' for which the $z=1$ is the usual Carrollian conformal algebra. The case $z=1/2$ occurs only in $d=1$ which is called Carroll-Schr\"odinger algebra \cite{Najafizadeh:2024imn}.  It would be interesting to study representation of these algebras and calculate their characters.

\paragraph{Conformal Carroll bootstrap.}
Although constructing Carrollian field theory actions can be very useful, one can tackle the  problem of constraing the parameter space of these theories by studying the correlations functions in the Carrollian theory at a conformal fixed point using a Carrollian conformal bootstrap program. The systematic first step towards this study is presented here for the two and three point functions.
One of the  features in Carrollian field theories is the emergence of electric and magnetic sectors \cite{Duval:2014uoa}. The electric sector theory is ultralocal as it is expected by the fact that the lightcone is closed. The magnetic sector describes a purely spatial theory whose energy is zero. We clearly see this distinction in calculating the two and three point functions. In the case of three-point function, we found an extra  possibility where the three points are forced to be aligned but not necessarily coincident. We think this term also belongs to the electric sector. This possibility is not invariant under temporal special conformal transformations. However, it seems that the SCT invariance is compatible with either the ultra-local theory or purely spatial theory.

\paragraph{Supersymmetry.}

Supersymmetric extensions of the Carroll algebra have been studied in the literature using the ultra-relativistic contraction method applied to relativistic algebras. These include the super-Carroll, super-AdS-Carroll, and Carrollian super-conformal algebras, as derived in \cite{Koutrolikos:2023evq, Bergshoeff:2015wma, Bagchi:2022owq}, respectively. Notably, the latter can be identified as the supersymmetric extension of the Carrollian conformal algebra (with $z=1$). Supersymmetric (holographic) examples of the $z=0$ case have  been studied in \cite{Afshar:2022mkf,Rosso:2022tsv}. Therefore, it would be of interest to study supersymmetric extensions of the different types of conformal Carroll algebras for general $z$.

\paragraph{Field theory.} 
From a field theory perspective, massive Carrollian scalar fields have been explored in works such as \cite{Henneaux:2021yzg,deBoer:2021jej,Bergshoeff:2022qkx}. Studies on Carrollian fermionic fields can be found in \cite{Bagchi:2022eui,Koutrolikos:2023evq,Banerjee:2022ocj,Bergshoeff:2023vfd}, while supersymmetric Carrollian fields have been examined in \cite{Koutrolikos:2023evq,Zorba:2024jxb}. These theories are characterized by the Carroll algebra and the super-Carroll algebra as their symmetries. However, field theories that exhibit different types of conformal Carroll algebras as their symmetry are relatively rare. Notable examples include theories where the Carrollian conformal algebra \eqref{cca} is the underlying symmetry \cite{Baiguera:2022lsw,Rivera-Betancour:2022lkc,Bekaert:2022oeh,Bagchi:2022eav}. Additionally, the Carroll-Schr\"odinger field theory, as presented in \cite{Najafizadeh:2024imn}, is governed by the Carroll-Schr\"odinger algebra \eqref{c-sch-a}. Interestingly, other conformal Carroll algebras identified in this work (i.e. type K, type D, type D-K) with arbitrary dynamical exponent $z$ do not yet have corresponding field theories, presenting an intriguing area for further research.

\section*{Acknowledgments}

MN is grateful to the members of the IPM HEP-TH weekly group meeting for discussions on related topics. The work of HA and MN is based upon research funded by Iran National Science Foundation (INSF) under project No. 4028530. MN is also partially supported by IPM funds.


\bibliographystyle{hephys}

\end{document}